\pdfoutput=1
\documentclass[sigconf]{acmart}
\usepackage{enumitem}
\usepackage{graphicx}
\usepackage{epstopdf}
\usepackage{amssymb}
\usepackage{amsmath,amsthm}
\usepackage{comment}
\usepackage{longtable}
\usepackage{multirow}
\usepackage{booktabs}
\usepackage{tabularx}
\usepackage{lineno}
\theoremstyle{definition}
\usepackage{balance}

\usepackage[font={small}]{caption}
\usepackage{setspace}

\usepackage{etoolbox}
\usepackage{tikzpagenodes}

\DeclareRobustCommand*\circled[1]{\tikz[baseline=(char.base)]{ \node[shape=circle,draw,color=white,fill=black,inner sep=0.5pt] (char){#1};}}
\usepackage{microtype}
\usepackage[labelsep=quad,indention=10pt]{subfig}
\usepackage{tikz}
\usepackage{pifont}
\newcommand{\cmark}{\ding{51}}%
\newcommand{\xmark}{\ding{55}}%
\newcommand{\pie}[1]{
  \begin{tikzpicture}
  \draw (0,0) circle (1ex);\fill (1ex,0) arc (0:#1:1ex) -- (0,0) -- cycle;
  \end{tikzpicture}
}
\usepackage[flushleft]{threeparttable}
\usepackage[ruled,vlined,linesnumbered]{algorithm2e}
\everypar{\looseness=-1}

\newcommand\numberthis{\addtocounter{equation}{1}\tag{\theequation}}
\def\ie{{i.e.},~}
\def\eg{{e.g.},~}

\newcommand{\argmin}{\operatornamewithlimits{argmin}}
\newcommand{\argmax}{\operatornamewithlimits{argmax}}
\renewcommand\footnotetextcopyrightpermission[1]{}

\begin{document}
\fancyhead{}

\title{Detection under Privileged Information (Full Paper)}
\titlenote{\small{A short version of this paper appears in ACM Asia Conference on Computer and Communications Security (ASIACCS) 2018.}}

\author{Z. Berkay Celik}
\affiliation{Pennsylvania State University}
\email{zbc102@cse.psu.edu}

\author{Patrick McDaniel}
\affiliation{Pennsylvania State University}
\email{mcdaniel@cse.psu.edu}

\author{Rauf Izmailov}
\affiliation{Vencore Labs}
\email{rizmailov@appcomsci.com}

\author{Nicolas Papernot, Ryan \nolinebreak Sheatsley, Raquel Alvarez}
\affiliation{Pennsylvania State University}
\email{{ngp5056, rms5643,rva5120}@cse.psu.edu}

\author{Ananthram Swami}
\affiliation{Army Research Laboratory}
\email{ananthram.swami.civ@mail.mil}

\mathchardef\UrlBreakPenalty=10000

\renewcommand{\shortauthors}{Z. Celik et al.}

\begin{abstract}
For well over a quarter century, detection systems have been driven by models learned from input features collected from real or simulated environments. An artifact (e.g., network event, potential malware sample, suspicious email) is deemed malicious or non-malicious based on its similarity to the learned model at runtime.  However, the training of the models has been historically limited to only those features available at runtime. In this paper, we consider an alternate  learning approach that trains models using ``privileged'' information--features available at training time but not at runtime--to improve the accuracy and resilience of detection systems. In particular, we adapt and extend recent advances in knowledge transfer, model influence, and distillation to enable the use of forensic or other data unavailable at runtime in a range of security domains. An empirical evaluation shows that privileged information increases precision and recall over a system with no privileged information: we observe up to 7.7\% relative decrease in detection error for fast-flux bot detection, 8.6\% for malware traffic detection, 7.3\% for malware classification, and 16.9\% for face recognition. We explore the limitations and applications of different privileged information techniques in detection systems. Such techniques provide a new means for detection systems to learn from data that would otherwise not be available at runtime.
\end{abstract}

\keywords{Detection systems; privileged information; machine learning}
  
\maketitle

\begin{tikzpicture}[remember picture,overlay]
    \node[align=center] at ([yshift=1em]current page text area.north) {A short version of this paper is accepted to ACM Asia Conference on Computer and Communications Security (ASIACCS) 2018 };
  \end{tikzpicture}%
  \vspace{-0.2in}

\section{Introduction}
Detection systems based on machine learning are an essential tool for system and enterprise defense~\cite{scarfone2007guide}. Such systems provide predictions about the existence of an attack in a target domain using information collected in real-time. The detection system uses this information to compare the runtime environment against known normal or anomalous states. In this way, the detection system ``recognizes'' when the environmental state becomes---at least probabilistically---dangerous. What constitutes dangerous is learned; detection algorithms construct models of attacks (or non-attacks) from past observations using a training algorithm. Thereafter, the detection systems use that model for detection at runtime. A limitation of this traditional approach  is that it relies solely on the features (also referred to as inputs) that are available at runtime.  In practice, many features are too expensive to collect in real-time or only available after the fact--and are thus ignored for the purposes of detection. 

Consider a recent event that occurred in the United States. In the Summer of 2017, the credit reporting agency Equifax fell victim to sophisticated cyber attacks that resulted in substantial exfiltration of personal information and intellectual property~\cite{Equifax}. Working with the government staff and security analysts conducted a clandestine investigation. During that time, a vast amount of information was collected from networks and systems across the agency, \eg network flows, system logs files and user activity. An analysis of the collected data revealed the presence of previously undetected advanced persistent threat (APT) actors on the agency's network. Yet, the collected analysis is largely non-actionable by detection systems post investigation; because the vast array of derived features would not be available at runtime, they cannot be used to train Equifax's detection systems.

In other contexts, features may be available at runtime but infeasible or undesirable to collect because of environmental or system constraints.  For example, the collection of a large numbers of features in environments of mobile phones~\cite{bickford2011security}, Internet of Things (IoT)~\cite{raza2013svelte}, sensor networks~\cite{hassanzadeh2013pride}, embedded control systems~\cite{reeves2011lightweight}, or ad-hoc networks~\cite{butun2014survey} is often too slow or requires too many resources to be feasible in practice. 

These examples highlight a challenge for future intrusion detection: \emph{how can detection systems integrate intelligence relevant to an attack that is not available at runtime?} Here, we turn to recent advances in machine learning that support models that learn on a superset of features used at runtime~\cite{vapnik2015learning,vapnik2009new}. The goal of the work described in this paper is to leverage these additional features, called {\it privileged information} (features available at training time, but not at runtime), to improve the accuracy of detection. Using this approach designers and operators of detection systems can leverage additional effort during system calibration to improve detection models without inducing additional runtime costs.

Pioneered recently by Vapnik, Izmailov, and others, learning under privileged information eliminates the need for symmetric features in training and runtime--thereby expanding the space of learning models to include ``ancillary" and ``non-runtime" information. However, to date, the application of these techniques in practical domains has been limited, and within the context of security non-existent. In this work, we explore how this new mode of learning can be leveraged in detection systems. This requires an exploration of not only use of these new learning models but also their applicability to security domains and the requirements of those domains on feature engineering.   Our experience in this effort over the last two years has demonstrated that blind application of privileged application can lead to poor detection--yet judicious and careful use can substantially improve detection quality. 

More concretely, in this paper, we explore an alternate approach to training intrusion detection systems that exploit privileged information. We design algorithms for three classes of privileged-augmented detection systems that use: (1) \emph{Knowledge transfer}, a general technique of extracting knowledge from privileged information by estimation from available information, (2) \emph{Model influence}, a model-dependent approach of influencing the model optimization with additional knowledge obtained from privileged information; and (3) \emph{Distillation}, an approach of summarizing the additional knowledge about privileged samples as class probability vectors. We further explore feature engineering in a privileged setting. To this end, we measure the potential impacts of privileged features on runtime models. Here, we use the degree to which a feature improves a model (\emph{accuracy gain}---a feature's additive contribution to accuracy) as a quality metric for selecting privileged features for a target model. We develop an algorithm and system that selects features that maximize model accuracy in the presence of privileged information. Finally, we compare the performance of privileged-augmented systems with systems without privileged information.  We evaluate four recently proposed detection systems: (1)  face authentication, (2) fast-flux bot detection, (3) malware traffic detection, and (4) malware classification. Our contributions are:

\vspace{-2pt}
\begin{itemize}
\item We augment several diverse detection systems using three classes of privileged information techniques and explore the strength and weaknesses of these techniques.

\item We present the first methods for feature engineering in privileged-augmented detection for security domains and identify inherent tensions between information utilization, detection accuracy, and model robustness.

\item We provide an evaluation of techniques on a variety of existing detection systems on real-world data. We show that privileged information decreases the relative detection error of traditional detection systems up to 16.9\% for face authentication, 7.7\% for fast-flux bot detection, 8.6\% for malware traffic detection, and 7.3\% for malware classification.

\item We analyze dataset properties and algorithm parameters that maximize detection gain, and present guidelines and cautions for the use of privileged information in realistic deployment environments.
\end{itemize} 

\noindent After introducing the technical approach for detection in the next section, we consider several key questions:
\begin{enumerate}
    \item How can the best features for a specific detection task be identified? (Section~\ref{sec:select-privileged})
    \item How does privileged-base detection perform against traditional systems? (Section~\ref{sec:experimental-results})
    \item How can we select the best privileged algorithm for a given domain and detection task? (Section~\ref{sec:selectionApproach})
    \item What are the practical concerns in using privileged information for detection? (Section~\ref{sec:discussion})
\end{enumerate}

\begin{figure}[t!]
\centering
\includegraphics[width=1\columnwidth]{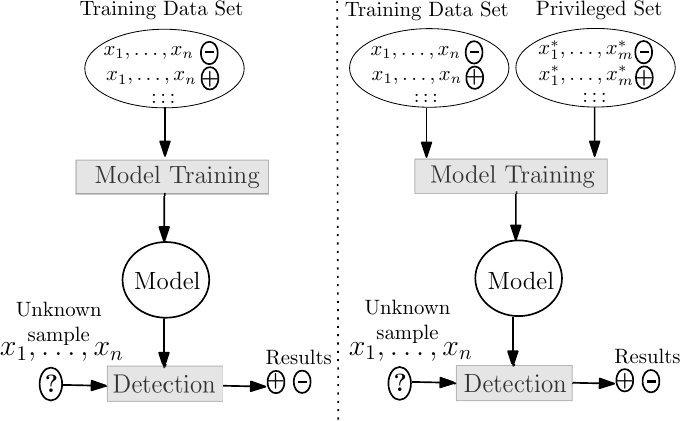}
\caption{Overview of a typical detection system (left) and proposed solution (right): Given training data including both malicious (+) and benign samples (--),  modern detection systems use training data to construct a model. To predict the class of an unknown sample, the system examines its features using the model to predict the sample as benign or malicious. In contrast, we construct a model both using training and privileged data, yet the model only requires the training data features to detect unknown sample as benign or malicious.}
\label{fig:motivation}
\end{figure}

\begin{figure*}[t!]
        \centering
        \resizebox{\textwidth}{!}{
           \subfloat[\normalsize{Knowledge transfer}]{%
              \includegraphics[height=6.5cm]{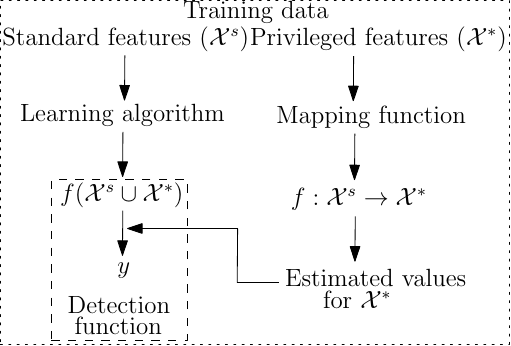}%
              \label{fig:approach-kt}%
           } 
           \hspace{0.05cm}
           \subfloat[\normalsize{Model influence}]{%
              \includegraphics[height=6.5cm]{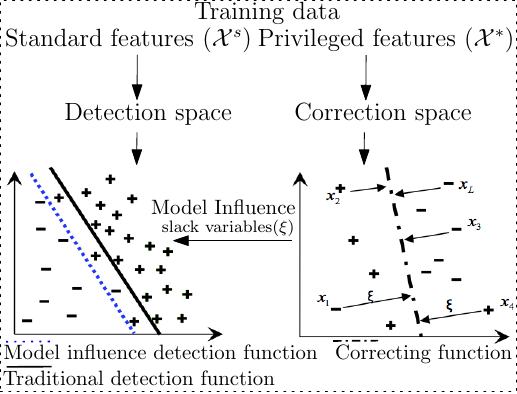}%
              \label{fig:approach-model-influence}%
           }
            \hspace{0.05cm}
           \subfloat[\normalsize{Distillation}]{%
              \includegraphics[height=6.5cm]{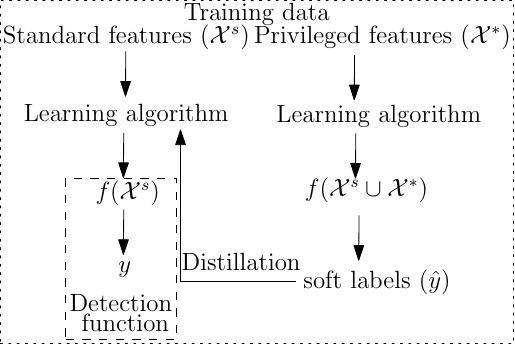}%
              \label{fig:approach-distillation}%
           } 
           }
           \caption{Schematic overview of approaches. Instead of making a detection model solely rely on standard features of a detection system, we also integrate privileged features into detection algorithms. }
           \label{fig:approaches}
 \end{figure*}

\section{Problem Statement} 
\label{sec:problem}
Detection systems use traditional learning algorithms such as support vector machines or multilayer perceptrons (neural networks) to construct \emph{detection models}. These models aim at learning patterns from historical data (also referred to as training data) to estimate an underlying dependency, structure or behavior of a system, process or environment. This \emph{training data} is a collection of samples that includes a vector of \emph{features} (e.g., packets per second, port number) and class \emph{labels} (e.g., anomalous or normal).   Once trained, runtime events (e.g., network event) are compared to the learned model.  Without loss of generality, the model outputs a label (or label confidence) that most closely fits with those of the training data.  The percentage of output labels that are correctly predicted for a sample set is known as its accuracy.

The quality of the detection system largely depends on features used to train models. In turn, the success of detection depends on explanatory variation behind the features that are used to separate an attack and benign sample. However, modern detection systems by construction assume that the features used to make predictions at runtime would be identical to those used for training (See Figure~\ref{fig:motivation} (left)). This assumption restricts the model training to the features that are available at runtime to make predictions. As highlighted above, intelligence obtained from forensic investigations~\cite{walls2011forensic}, data obtained through a human expert analysis~\cite{cardenas2013big}, or features unable to be feasibly collected at runtime is simply not actionable.  Juxtapose this with our goal of leveraging features in model training that are not available at runtime to improve detection accuracy (See Figure~\ref{fig:motivation} (right)). Note that we do not focus on a specific detection task or domain.  We begin in the following section by introducing the three approaches we use to integrate privileged information into detection models.

\section{Privileged Information Detection}
\label {sec:methodology}
This section introduces three approaches to integrate privileged features into detection algorithms: knowledge transfer, model influence, and distillation. Figure~\ref{fig:approaches} presents the schematic overview of approaches. Stated formally, we consider a conventional detection algorithm as a function $f : \mathcal{X} \to \mathcal{Y}$ that aims at predicting \emph{targets} $y \in \mathcal{Y}$ given some explanatory \emph{features} $x \in \mathcal{X}$. The models are built using a \emph{dataset} containing pairs of features and targets, denoted by $\mathcal{D} = \{(x_i, y_i)\}_{i=1}^n$. Following the definition of privileged information, we consider a detection setting where the features $\mathcal{X}$ used for detection are split into two \emph{sets} to characterize their availability for detection at training time. The \emph{standard features} $\mathcal{X}^{s} = \{{x}_i, i=1,\ldots, n, {x}_i \in \mathbb{R}^d\}$ includes the features that are reliably available both at training and runtime (as in conventional systems), while \emph{privileged features} $\mathcal{X^*} = \{{x}^*_i, i=1,\ldots, n, {x}^*_i \in \mathbb{R}^m\}$ have constraints that prevent using them for detection at runtime. More formally, we assume that detection models will be trained on some data $\{(x_i, x^*_i, y_i)\}_{i=1}^n$, and they will make detection on some data $\{x_j\}_{j=n+1}^{n+m}$. Therefore, our goal is creation of algorithms that efficiently integrate privileged features $\{x^*_i\}_{i=1}^n$ into detection models, without requiring them at runtime. 

\subsection{Knowledge Transfer} 
\label{sec:knowledgeTransfer}
We consider a general algorithm to transfer knowledge from the space of privileged information to the space where the detection model is constructed~\cite{vapnik2015learning}. The algorithm works by deriving a \emph{mapping function} to estimate each privileged feature from a subset of standard features. The algorithm used to identify a mapping function  $f_i$ is described in Algorithm~\ref{algo:knowledgeTransfer}. The estimation is straightforward: the relationship between standard and privileged features is learned by defining each privileged feature $x_i^*$ as a target and standard set as an input to a mapping function $f_i \in \mathcal{F}$ (lines 1-3). The mapping functions can be defined in the form of any function such as regression or similarity-based (we give examples in Section~\ref{sec:evaluation-kt}). The use of the mapping function allows a system to apply the same model learned from the complete set at training time with the union of standard and estimated privileged features on unknown samples (See Figure~\ref{fig:approach-kt}).  By using $f_i$, detection systems are able to construct the complete features with the correct values of privileged features at runtime--intuitively, each $f_i$ is used to generate a synthetic feature that represents an estimate of a privileged feature (line 4). As a result, the accurate estimation of privileged features contributes to using complete and relevant features in a model training and, therefore, enhances the generalization of models compared to those trained solely on standard features. Note that the estimating power of $f_i$ is bounded by the size and completeness of the training data (with respect to the privileged features), and thus the use of $f_i$ in the model should be calibrated based on measurements of estimation quality (See Section~\ref{sec:model-comparison} for details).

\begin{algorithm}[t!]
\setstretch{0.8}
\SetKwInOut{Input}{Input}\SetKwInOut{Output}{Output}
\Input{Standard training set $\mathcal{X}^{s} = \vec{x_1},\ldots,\vec{x_L}, \vec{x_i} \in \mathbb{R}^d$\newline
Privileged training set $\mathcal{X}^{*}=\vec{x_1^*},\ldots,\vec{x_L^*}, \vec{x_i^*} \in \mathbb{R}^m$\\
$f_i \in F \gets$ mapping function\\
$\theta \gets$ mapping function parameters}
Find an optimal standard set $\widehat{\mathcal{X}} \subseteq \mathcal{X}^s$, for all $x^{*}_{i} \in \mathcal{X}^*$\\
Select set $\widehat{\mathcal{X}}_i$ for $x_i^*$ \\
Mapping function $f$ evaluates $(\widehat{\mathcal{X}}_i,x_{i}^{*})$ as given $x_{i}^{*}= f_i(\widehat{\mathcal{X}}_i,\beta | \theta)$, $f_i \in\mathcal{F}_m$\\
Output $\beta$ for $x_i^*$ that minimizes $f_i$\\
At runtime use $f_i$ to estimate $x_i^*$ 
\caption{Knowledge Transfer Algorithm}
\label{algo:knowledgeTransfer}
\end{algorithm}

\subsection{Model Influence}
\label{sec:modelInfluence}
Model influence incorporates the useful information obtained from privileged features to the correction space of the detection model by defining additional constraints on the training errors (See Figure~\ref{fig:approach-model-influence})~\cite{vapnik2009new, vapnik2015learning}. Intuitively, the algorithm learns how privileged information influences outputs on training input feature vectors towards building a set of corrections for the space of inputs--in essence creating a correction function that takes as input runtime features and adjusts model outputs. Note that while we adapt model influence  to support vector machines (SVM) herein, it is applicable to other ML techniques. More formally, consider a training datum that is generated from an unknown probability distribution $p(\mathcal{X}^{s},\mathcal{X}^{*},y)$. Our goal is to find a model $f : {\mathcal{X}}^{s} \rightarrow \mathcal{Y}$ such that the expected loss is defined as:
\begin{equation}
 R(\theta)= \int_{\mathcal{X}^{s}} \! \int_\mathcal{Y} \mathcal{L}((x^{s},\theta),y) p({x}^{s}, {x}^{*}, y)\,d {x}^{s}\,dy
\label{eq:lupi}
\end{equation}
\noindent Here the system is trained on standard and privileged features but only uses standard features at runtime. We consider the optimization problem of SVM in its dual form as shown in Equation~\ref{svmplusMain} (labeled as primary detection objective) where $\alpha$ is Lagrange multiplier. 
\begin{align*}
\mathcal {L} (w,w^*,b, b^*,\alpha, \delta) &= \overbrace{\frac{1}{2}||w||^2 + \sum_{i=1}^{m}\alpha_{i} -\sum_{i=1}^{m}\alpha_{i}y_if_i}^\textrm{main detection objective} \\[-0.3em]
& + \underbrace{\frac{\gamma}{2}||w^*||^2+\sum_{i=1}^{m}(\alpha_i+\delta_i -C)f_{i}^{*}}_\textrm{influence from privileged features}\numberthis
\label{svmplusMain}
\end{align*}

We influence the detection boundary of a model trained on standard features $f_i=w^\intercal x_{i}^{s} + b$ at $x_{i}^{s}$ with the correction function defined by the privileged features $f_i^{*} = {w^{*}}^\intercal x_{i}^{*}+ b^*$ at the same location (labeled as influence from privileged features). In this manner, we use privileged features as a correction function of the slack variables $\xi_i$ defined in the objective of SVM. In turn, the useful information obtained from privileged features is incorporated as a measure of confidence for each labeled standard sample. The formulation is named as SVM+ and requires $O(\sqrt n)$ samples to converge compared to $O(n)$ samples for SVM which is useful for systems with a sparse data collection~\cite{techReportLupi,vapnik2015learning}. We refer readers to Appendix~\ref{sec:appendixA} for a complete formulation and implementation. 

\begin{figure}[t!]
\centering
\includegraphics[width=0.9\columnwidth]{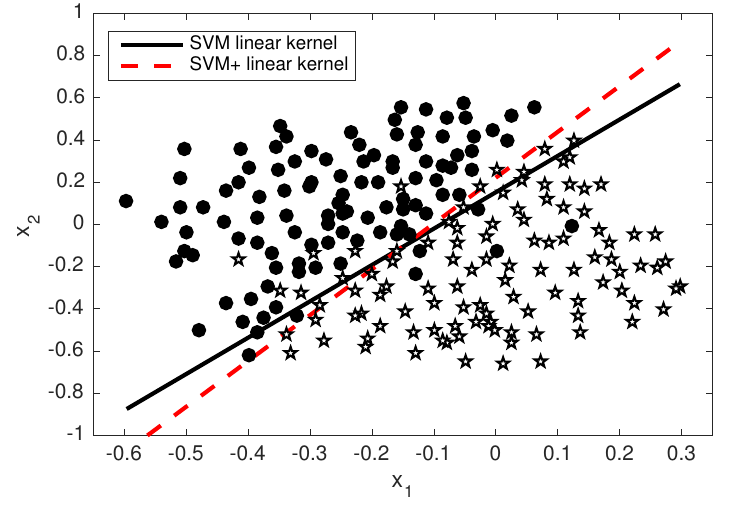}
\caption{Model influence: The SVM+ uses privileged information to correct the decision boundary of the model.}
\label{fig:SVMPlus1}
\end{figure}

To illustrate, consider the 2-dimensional synthetic dataset presented in Figure~\ref{fig:SVMPlus1}, as well as the decision boundaries of two detection algorithms SVM (an unmodified Support Vector Machines) and SVM+ (the same SVM augmented with model influence correction). The use of privileged information in model training separates the classes more accurately because privileged features accurately transfer information to standard space, and the resulting model becomes more robust to the outliers. This approach may provide even better class separation in datasets with higher dimensionality.

To summarize,  as opposed to knowledge transfer, we eliminate the task of finding the mapping functions between standard and privileged features. Thus, we reduce the problem of model training to a single unified task.

\begin{algorithm}[t!]
\setstretch{0.8}
\SetKwInOut{Input}{Input}\SetKwInOut{Output}{Output}
\Input{Standard training set $\mathcal{X}^{s} = \vec{x_1},\ldots,\vec{x_L}, \vec{x_i} \in \mathbb{R}^d$\newline
Privileged training set $\mathcal{X}^{*}=\vec{x_1^*},\ldots,\vec{x_L^*}, \vec{x_i^*} \in \mathbb{R}^m$\newline
Training labels $\mathcal{Y}$\newline
$T$ $\gets$ Temperature parameter, $T>0$ \newline
$\lambda$ $\gets$ Imitation parameter, $\lambda \in $ [0,1] } 
Learn a model $f_{s}$  using  $(\mathcal{X^*}, y)$\\
Compute soft labels as $s_i= \sigma(f_{s}(x_{i}^{*})/T)$ for all $1\leq i \leq L$\\
Learn detection model using Equation~\ref{eq:distillation} with the given data as $\{(\mathcal{X}^{s}, \mathcal{Y}), (\mathcal{X}^s, \mathcal{S}) \}$\\
Make detection using $\theta$ minimizing $f_t$ for standard features $x^s$ 
\caption{Distillation Algorithm}
\label{algo:generalizedDistillation}
\end{algorithm}

\subsection{Distillation}
Model compression or distillation are techniques to transfer knowledge from a complex Deep Neural Network (DNN) to a smaller one without loss of accuracy~\cite{hinton2015distilling}. The motivation behind the idea suggested in~\cite{ba2014deep} is closely related to knowledge transfer. The goal of the distillation is to use the class knowledge from both class labels (\ie hard labels) and probability vectors of each (\ie soft labels). The benefit of using class probabilities in addition to the hard labels is intuitive because probabilities of each class define a similarity metric over the classes apart from the samples' correct classes. Lopez-Paz et al. recently introduced an extension of model distillation used to compress models built on a set of features into models built on a different set of features~\cite{lopez2015unifying}. We adapt this technique to detection algorithms.

We address the problem of privileged information using distillation as follows. First, we train a ``privileged'' model on the privileged set and labels whose output of the model is the vector of soft labels $S$. Second, we train a \emph{distilled} model (used at runtime) by minimizing Equation~\ref{eq:distillation}, which learns a detection model by simultaneously imitating the privileged predictions of the privileged model and learning the targets of the standard set. The algorithm for learning such a model is presented in Algorithm~\ref{algo:generalizedDistillation} and outlined as follows:

{\small{
\begin{align*}
f_t(\theta) &=  \argmin_{f \in F_t} \frac{1}{n} \sum_{i=1}^{n} \bigg(  \overbrace{(1-\lambda) \mathcal{L}(y_i, \sigma (f (x^{s}_{i}))}^\textrm{detection} + \overbrace{\lambda \mathcal{L}(s_i, \sigma (f (x^{s}_{i})) }^\textrm{imitate privileged set} \bigg)\numberthis
\label{eq:distillation}
\end{align*}}}

We learn a privileged model $f_s \in F_s$ by using the privileged samples available at training time (line 1). We then compute the soft labels by applying the softmax function (\ie normalized exponential) $s_{i} = \sigma(f_s (x^{*}_{i})/{T})$ (line 2). The output is a vector which assigns a probability to each class of the privileged samples. We note that class probabilities obtained from privileged model provide additional information for each class. Here, temperature parameter $T$ controls the degree of class prediction smoothness. Higher $T$ enables softer probabilities over classes and vice versa. As a final step, Equation~\ref{eq:distillation} is sequentially minimized to distill the knowledge transferred from privileged features as a form of probability vectors (soft labels) into the standard sample classes (hard labels) (line 3). In Equation~\ref{eq:distillation}, the $\lambda$ parameter controls the trade-off between privileged and standard features. For $\lambda \approx 0$, the objective approaches the standard set objective, which amounts to detection solely on standard features. However, as $\lambda \to 1$, the objective transfers the knowledge acquired by the privileged model into the resulting detection model. Therefore, learning from the privileged model does, in many cases, significantly improve the learning process of a detection model.

Distillation differs from model influence and knowledge transfer in at least two ways. First, while knowledge transfer attempts to estimate the privileged features with a representation of a mapping function, distillation is a trade-off between the privileged sample probabilities and standard sample class labels. Second, in contrast to model influence, distillation is independent of the machine learning algorithm (model-free), and its objective function can be minimized using a model of choice.

\section{Privileged Information Systems}
\label{sec:select-privileged}
In this section, we explore algorithms for feature engineering (selecting privileged features for a detection task) and demonstrate their use in diverse experimental systems.

\begin{figure}[t!]
\centering
\includegraphics[width=0.8\columnwidth]{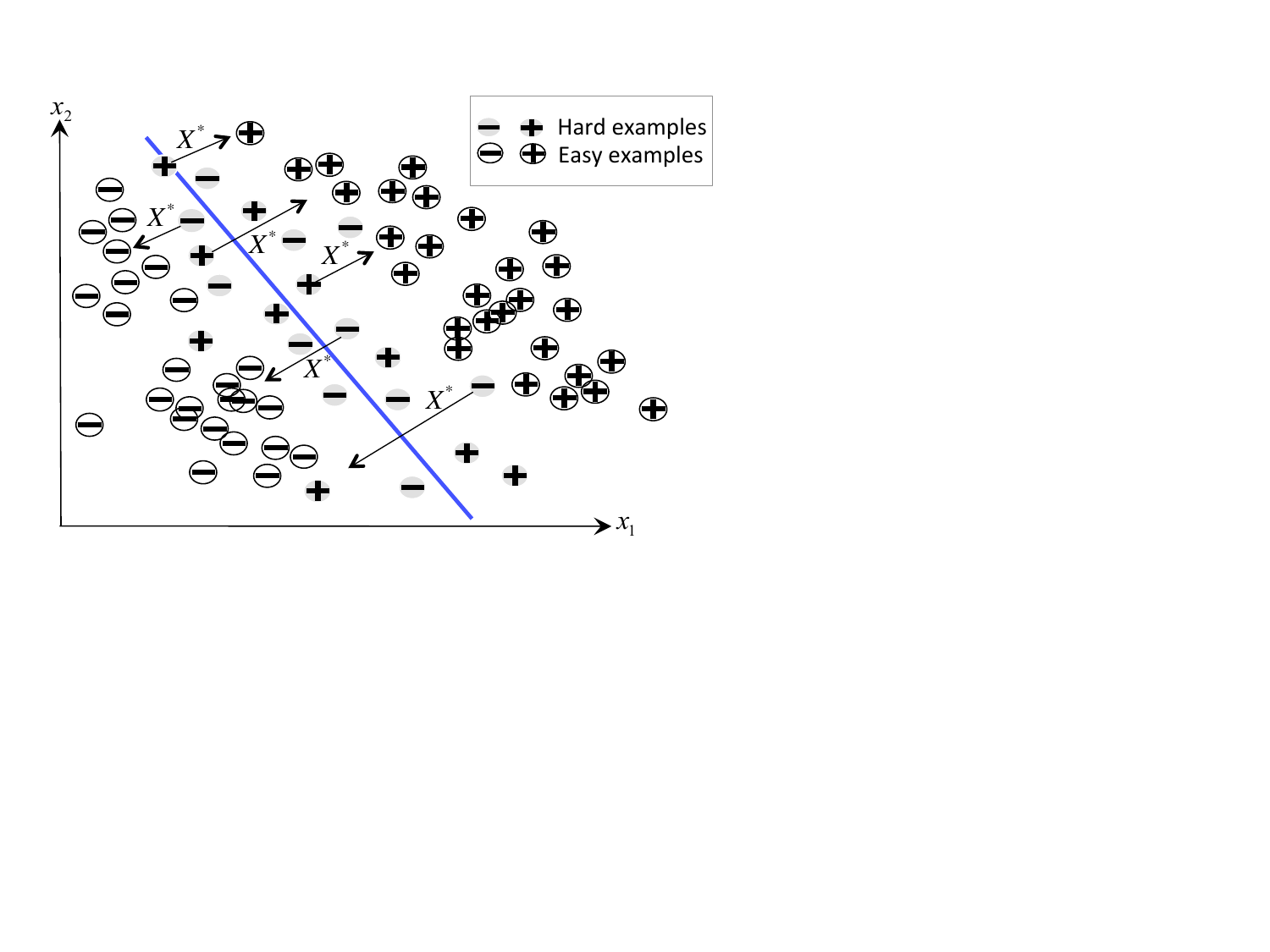}
\caption{Visualizing hard and easy benign (-) and malicious (+) examples. We select privileged features that aims at increasing the detection of hard examples.}
\label{fig:easy-hard-examples}
\end{figure}

\subsection{Selecting Privileged and Standard Features} 
The first challenge facing our model of detection is deciding which features should be used as privileged information.  Asked another way, \emph{given some potentially large universe of offline features, which are the most likely to improve detection?}  To address this, we develop an iterative algorithm that selects features that maximize model accuracy. Selection is made on the calculated \emph{accuracy gain} of each feature--a measure of the additive value of the features concerning an existing feature set for detection accuracy.

Our feature selection algorithm measures the potential impacts of privileged features that help detect the hard-to-classify examples. Generally speaking, easy examples fall in a distribution that can be explained by some set of model parameters, and hard examples do not precisely fit the model--and are either misclassified or near a decision boundary (See Figure~\ref{fig:easy-hard-examples})~\cite{sharmanska2013learning}. As a consequence, accurate classification of hard examples is one of the main challenges of practical systems, as they are the main source of detection errors due to the incorrect or insufficient information about normal or anomalous states of a system. 

The first step of feature engineering--as is true of any detection task--is identifying all of the available features that potentially may be used for detection. Specifically, we collect the set of domain-specific features based on using domain knowledge and surveying the recent efforts in that domain. It is from that set that we will identify the privileged features to be used for training.  Note that defining privileged features sometimes requires a level of domain expertise. However, trained security experts will find most privileged features straightforward after defining the runtime constraints on features. Additionally, the initial privileged set may include irrelevant features that carry little or no useful information for the target detection task; thus we identify the privileged set using Algorithm~\ref{algo:related-complete}. The algorithm starts with standard features of a detection system and sequentially adds one privileged feature from the set which maximizes correct classification of hard examples, i.e., the feature whose addition to the existing set has the greatest positive impact on accuracy (measured accuracy gain). The accuracy gain of hard examples is found using SVM classifier (model $J$ in algorithm~\ref{algo:related-complete}). This process is repeated until the potential feature set is empty, a maximum number of features is reached, or the accuracy gain is below a threshold for usefulness.  

\begin{algorithm}[t!]
\setstretch{0.8}
\SetKwInOut{Input}{Input}\SetKwInOut{Output}{Output}
\Input{Standard feature set   $x^s$ \newline
Related privileged feature set $x^{*}$\newline
SVM Detection model $J$ evaluating accuracy of hard-to-classify examples
}
\BlankLine 
Start with the standard set $Y_0=x^s$\\
Select the next privileged feature $x^+ = \argmax_{x^* \not\in Y_k} [\mathcal{J}(Y_k + x^*)]$\\
Update $Y_{k+1} = Y_k + x^+$; $k=k+1$\\
Go to 2\\
Output $Y$ including standard and selected privileged features
\caption{Selecting privileged features}
\label{algo:related-complete}
\end{algorithm}

\begin{table*}[th!]
\centering
\resizebox{\textwidth}{!}{%
\renewcommand{\arraystretch}{1.1}
\addtolength{\tabcolsep}{-4pt}
\begin{tabular}{cclll}
\hline
\multicolumn{1}{c}{\textbf{System}} & \multicolumn{1}{c}{\textbf{Datasets and}} & \multicolumn{1}{c}{\textbf{Standard features}} & \multicolumn{1}{c}{\textbf{Incorporated privileged features}} & \multicolumn{1}{c}{\textbf{Detection time constraints on privileged features}} \\\hline

{\normalsize{\circled{1}}} & \cite{wolf2011effective, fddbTech, LFWTechUpdate} & \multirow{2}{*}{\--Raw face images} & \multirow{2}{*}{\--Bounding boxes and cropped versions of facial images} & \--Need of additional software for processing \\
 &  &  &  &\--Infeasible in energy and processing constrained sensors\\\hline

\multirow{5}{*}{\normalsize{{\circled{2}}}} & \multirow{5}{*}{\cite{celik2013detection,huang2010fast,hsu2010fast,yadav2010detecting,passerini2008fluxor}} & \--Number of unique A and NS records in DNS packets  & \--Edit distance, KL divergence and Jaccard index (domain names) & \--Processing overhead of   whitelist domains\\
 &  & \--Network, processing and document fetch delay  & \--Time zone entropy of A and NS records in DNS packets & \multirow{2}{*}{\--IP coordinate database processing overhead} \\
 &  &  & \--Euclidean distance between server IP  and  NS address &  \\
 &  &  & \--Number of distinct Autonomous systems and networks & \--Time consuming WHOIS processing \\
 &  & \multicolumn{1}{c}{$\ldots$} & \multicolumn{1}{c}{$\ldots$} & \multicolumn{1}{c}{$\ldots$} \\ \hline
 
\multirow{5}{*}{\normalsize{\circled{3}}} & \multirow{5}{*}{\cite{miller2012sequential,celik2011salting, zou2011flow, berkay2015malware}} & \--Data bytes divided by the total number of packets & \--Source and destination port numbers &  \multirow{2}{*}{\--Adversary easily change them in subsequent malware versions}  \\
 &  & \--Total number of RTT samples found & \--Byte frequency distribution in packet payload &  \\
 &  & \--The count of all packets with at least a byte payload & \--Total connection time & \--Payload encryption in subsequent malware versions\\
 &  & \--The median of total IP packets & \--Total number of packets with URG and PUSH flag set &  \\
 &  & \multicolumn{1}{c}{$\ldots$} & \multicolumn{1}{c}{$\ldots$} & \multicolumn{1}{c}{$\ldots$} \\ \hline

\multirow{2}{*}{\normalsize{\circled{4}}} & \multirow{2}{*}{\cite{ahmadi2015novel,kaggleMicrosoft}} & \multirow{2}{*}{\--Frequency count of hexadecimal duos in binary files} & \multirow{2}{*}{\--Frequency count of distinct tokens in metadata log} & \--Software-dependency of obtaining assembly source code \\
 &  &  &  &\--Computational overhead and error-prone feature acquisition \\\hline
\end{tabular}%
}
\caption{Description of detection systems: {\footnotesize{\circled{1}}} Face Authentication, {\footnotesize{\circled{2}}} Fast-flux Bot Detection, {\footnotesize{\circled{3}}} Malware Traffic Detection, and {\footnotesize{\circled{4}}} Malware Classification. Appendix~\ref{sec:AppendixC} provides details on the standard and privileged features used throughout.}
\label{table:datasetDescriptions} 
\end{table*}

Note that the quality of the selection process is a consequence of the training data used to calculate accuracy gain.  If the training data is not representative of the runtime input distribution, the algorithm could inadvertently over or under-estimate the accuracy gain of a feature and thereby weaken the detection system~\cite{celik2017feature}.  Additionally, the accuracy gain can be computed via different feature selection algorithms. For instance, a more brute force approach that evaluates the impact of all standard and privileged feature combination can be used. This drastically increases computation overhead with the high dimensional datasets used in our evaluation. Note that this limitation is not unique to feature selection in this context, but applies to all feature engineering in extant detection systems.

\subsection{Experimental Systems}
\label{sec:exp-systems}
In this section, we introduce four security-relevant systems for face authentication, fast-flux bot detection, malware traffic detection, and malware classification. We selected these experimental systems based on their appropriateness and diversity of their detection task.  This diverse set of detection systems serves as a representative benchmark suite for our approaches. The following are the steps involved in constructing  each system (discussed below):
\begin{enumerate}
\item Extract standard features of existing detection systems 
\item Add new privileged features by identifying detection-time constraints on the features 
\item Use the algorithm in preceding section to calibrate the detection system with standard and privileged features
\end{enumerate}

Through this process, we construct their privileged-augmented systems with application of approaches that is used for the validation  in the following section. Table~\ref{table:datasetDescriptions} summarizes the experimental systems and the standard and privileged features selected. Additional details about these systems and their features are presented in Appendix~\ref{sec:AppendixC}.

\begin{figure}[t!]
\centering 
\includegraphics[width=0.8\columnwidth]{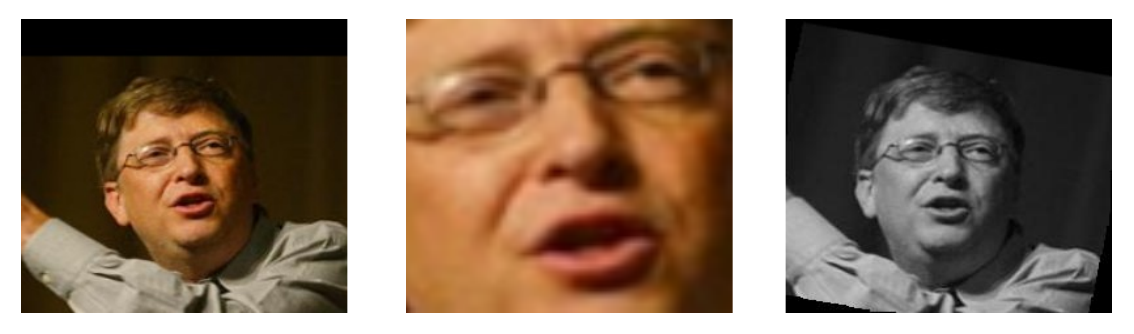}
\caption{Example of face authentication features. Original image for standard features (Left), cropped (Middle), and funneled (Right) images used for privileged features. }
\label{fig:billGates}
\end{figure}

\vspace{2pt}\noindent{\bf  Experimental System: Face Authentication} - 
To explore the efficiency of approaches in image domains, we modeled a user authentication system based on recognition of facial images. Our goal is to recognize an image containing a face with an identifier corresponding to the individual depicted in the image. We use images from a public dataset that includes face images labeled with each person's name~\cite{LFWTechUpdate}. We build the features from 1348 facial images with at least 50 images per user.

\vspace{2pt}\noindent\textit{Privileged information.} It is recently found that face recognition systems used for access control in particular energy and computation constrained camera sensors can be easily bypassed by an attacker~\cite{duc2009your}. In this, the lack of useful features or number of images used to train the systems is the main reason of duping the systems into falsely authenticate/recognize users. We use two types of privileged features for each image in addition to the original images in model training: cropped and funneled versions of the images (See Figure~\ref{fig:billGates})~\cite{wolf2011effective,fddbTech}. These images provide additional information for a users' face by image aligning and localizing~\cite{huang2012learning}. While it is technically possible these features can be obtained by an aid of software or human expert at runtime, they are much more likely to not be available in low energy and  slow processing sensors (and thus we define them as privileged).\footnote{We interpret the accuracy gain as a defense for hardening misclassification of users.}

\vspace{2pt}\noindent{\bf Experimental System: Fast-flux Bot Detection} - The Fast-flux bot detector is used to identify hosts that use fast-changing DNS entries to hide the existence of server hosts used for malicious activities. The raw data consists of 4GB DNS requests of benign and active fast-flux servers collected in early 2013~\cite{celik2013detection}. We build a detection system by using the 19 features used in recently proposed botnet detectors~\cite{yadav2010detecting,passerini2008fluxor,huang2010fast}. This system relies on features obtained from domain names, DNS packets, packet timing intervals, WHOIS domain lookup and IP coordinate database. The resulting dataset includes many features to increase separation of Content Delivery Networks (CDNs) from fast-flux servers, as similarities between them are the main source of detection errors. 

\vspace{2pt}\noindent\textit{Privileged information.} In this system, even though the complete features are relevant for fast-flux detection, obtaining some features at runtime entails computational delays. For example, processing of WHOIS records, maintaining up-to-date IP coordinate database and whitelist of domain names takes several minutes/hours. Thus, we define eleven features obtained from these sources as privileged to assure real-time detection.

\begin{figure}[t]
\small
\centering
\includegraphics[width=\columnwidth]{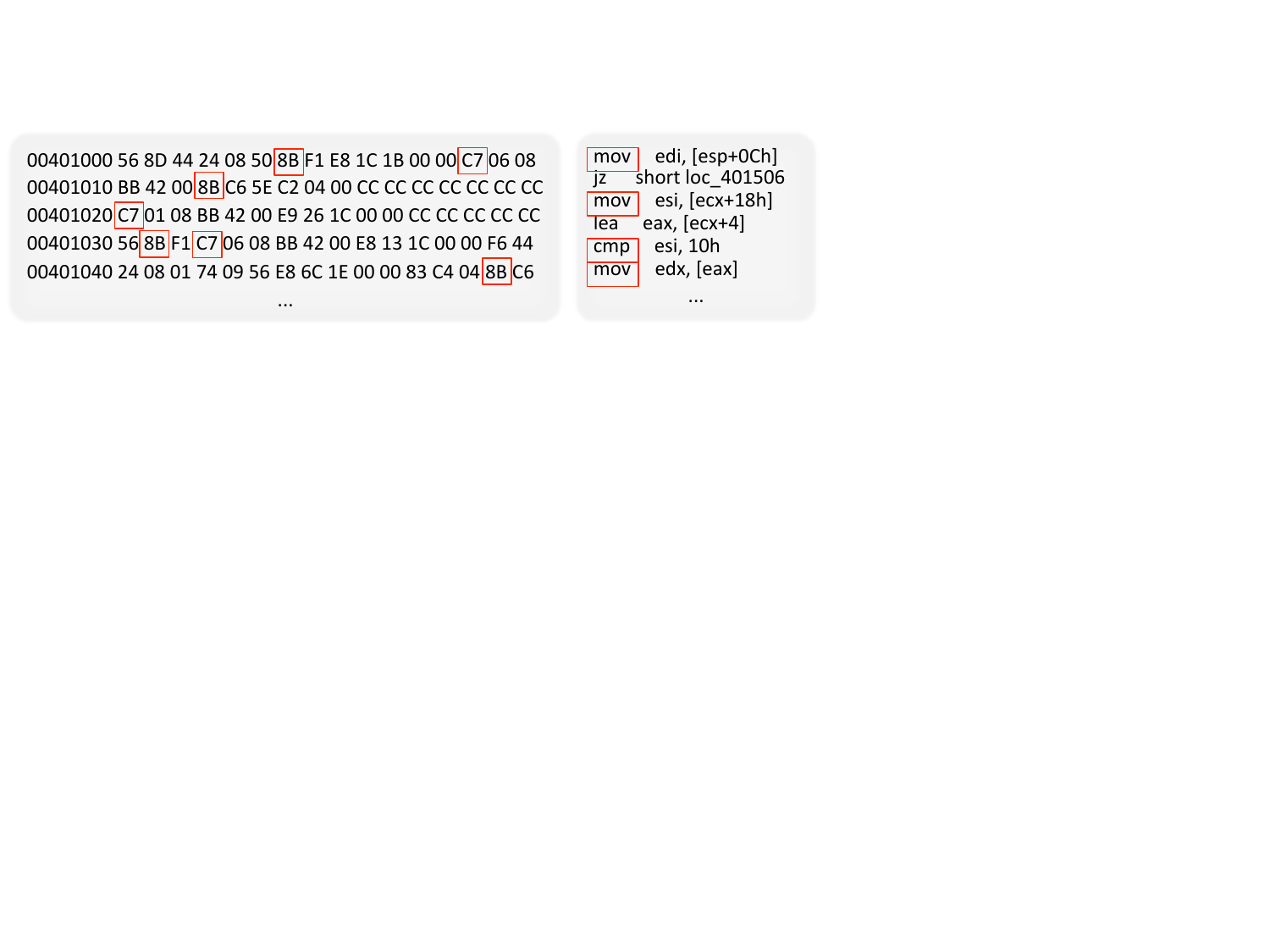}
\caption{Excerpt from hexadecimal representations (right), and assembly view (left) of an example malware. Selected byte bigrams and tokens for this malware is shown in boxes.}
\label{fig:malware-features}
\end{figure}

\begin{table*}[ht!]
\small
\centering
\resizebox{\textwidth}{!}{%
\addtolength{\tabcolsep}{3pt}
\renewcommand{\arraystretch}{1.1}
\begin{tabular}{c|c|ccc|c|c}
\hline
\multirow{2}{*}{\textbf{System}} & \multirow{2}{*}{\textbf{Approach}} & \multicolumn{3}{c|}{\textbf{Relative Gain over Traditional Detection}} & \multirow{2}{*}{\textbf{\begin{tabular}[c]{@{}c@{}}Parameter \\ Optimization?\end{tabular}}} & \multirow{2}{*}{\textbf{\begin{tabular}[c]{@{}c@{}}Detection time \\ Overhead?\end{tabular}}} \\ \cline{3-5}
 &  & \textbf{Accuracy} & \textbf{Precision} & \textbf{Recall} &  &  \\ \hline
Fast-flux Bot Detection & \begin{tabular}[c]{@{}c@{}}Knowledge Transfer \end{tabular} & 7.7\% & 5.3\% & 3.4\% & \cmark & \cmark \\
Malware Classification & Model Influence & 7.3\% & 2.2\% & 3.7\% & \cmark & \xmark \\
Malware Traffic Analysis & Distillation & 8.6\% & 2.2\% & 5\% & \cmark & \xmark \\
Face Authentication & Distillation & 16.9\% & 9.3\% & 9.2\% & \cmark & \xmark \\ \hline
\end{tabular}%
}
\caption{Summary of accuracy, precision and recall gain via privileged information-augmented systems. The best resulting approach for each detection system is illustrated. (See the next two sections for more details on approach comparison and runtime overhead of approaches.)}
\label{summary-results}
\end{table*}

\vspace{2pt}\noindent{\bf  Experimental System: Malware Traffic Detection} - Next, we modeled a malware traffic anomaly detection system based on network flow statistics used in recent detection systems~\cite{gu2008botminer, miller2012sequential,celik2011salting}. The system aggregates 20 flow features for detecting botnet command and control (C\&C) activity among benign applications. For instance, the ratio between maximum and minimum packet size from server to client and client to server find out to be a distinctive observation between benign and malicious samples. We add 173 botnet traffic of Zeus variants that is used for spam distribution, DDoS attacks, and click fraud~\cite{rossow2013sok, garcia2014empirical} to the 1553 benign applications (web browsing, chat, email, etc..) of Lawrence Berkeley National Laboratory (LBNL)~\cite{pang2005first} and University of Twente~\cite{barbosa2010simpleweb}.

\vspace{2pt}\noindent\textit{Privileged information.} In this system, the authors eliminate the features that can be readily altered by an attacker, as the model trained with tampered features allows an attacker to manipulate the detection results easily~\cite{zou2011flow, mcdaniel2016machine}. The impact of altering the features at run time on detection accuracy is recently studied~\cite{celik2011salting}. For instance, consider that destination port numbers or packet interval time are used as a feature. An adversary may ``easily" change them in subsequent malware versions to evade detection systems. Also, the authors do not use payload content to obtain features because the attacker can use encrypted traffic to prevent deep packet inspection. Thus, we deem such eight features as a privileged, as inference does not consider their tampered values at runtime.

\vspace{2pt}\noindent{\bf  Experimental System: Malware Classification} - The Microsoft malware dataset~\cite{kaggleMicrosoft} is an up-to-date publicly available corpus. The dataset includes nine malware classes including hexadecimal representation of the malware's binary content, and a class representing one of nine family names. The dataset used in our experiments includes 1746 malware samples extracted from 200GB malware files. We build a real-time malware classification system by using the binary content file. Following a recent malware classification system~\cite{ahmadi2015novel}, we construct features by counting frequencies of each hexadecimal duos (\ie byte bigrams). These features found out to provide distinctive between different families because of exploiting the code dissimilarities among families. 

\vspace{2pt}\noindent\textit{Privileged information.} This dataset also includes a metadata manifest log file. The log file contains information such as memory allocation, function calls, strings, etc.. The logs along with the malware files can be used for classifying malware into their respective families. Thus, similar to the byte files, we obtain the frequency count of distinct tokens from asm files such as $\texttt{mov()}$, $\texttt{cmp()}$ in the text section (See Figure~\ref{fig:malware-features}). These tokens allow us to capture execution differences between different families~\cite{ahmadi2015novel}. However, in practice, obtaining features from log files introduces significant overheads in the disassembly process. Further, various types or versions of a disassembler may output byte sequences differently. Thus, this process may result in inaccurate and slow feature processing in real-time automated systems~\cite{nataraj2011comparative}. To address these limitations, we include features from disassembler output as privileged for accurate and fast classification.

\begin{table}[t!]
\small
\centering
\renewcommand{\arraystretch}{1}
\addtolength{\tabcolsep}{-3pt}
\resizebox{\columnwidth}{!}{
\begin{tabular}{@{}lccc@{}}
\toprule
                                   & \textbf{{\scriptsize{Knowledge Transfer}}}              & \textbf{{\scriptsize{Model Influence}}}                 & \textbf{{\scriptsize{Distillation}}                     } \\ \midrule
\textbf{Fast-flux Bot Detection}    & Section~\ref{sec:evaluation-kt}    & Section~\ref{sec:model-comparison} & Section~\ref{sec:model-comparison} \\
\textbf{Malware Traffic Detection} & Section~\ref{sec:model-comparison} & Section~\ref{sec:ev-model-influence} & Section~\ref{sec:model-comparison} \\
\textbf{Face Authentication}       & ---                                       & ---                                       & Section~\ref{sec:ev-distillation}  \\
\textbf{Malware Classification}    & Section~\ref{sec:model-comparison} & Section~\ref{sec:model-comparison} & Section~\ref{sec:model-comparison} \\ \bottomrule
\end{tabular}
}
\caption{Summary of validation experiments.}
\label{table:roadmap}
\end{table}

\section{Evaluation}
\label{sec:experimental-results}
In this section, we explore the following questions:

\begin{enumerate}[leftmargin=*]
\item \emph{How much does privileged-augmented detection improve performance over systems with no privileged information?} We evaluate the accuracy, precision, and recall of approaches, and demonstrate the detection gain of including privileged features.

\item \emph{How well do the approaches perform for a given domain and detection task?} We answer this question by comparing the results of approaches and present guidelines and cautions for appropriate approach calibration to maximize the detection gain.

\item \emph{Do approaches introduce training and detection overhead?} We report model learning and runtime overhead of approaches for realistic environments.  
\end{enumerate}

Table~\ref{summary-results} shows the summary of the privileged information augmented systems, and Table~\ref{table:roadmap} identifies the validation experiments described throughout. As detailed throughout, we find that the use of privileged information can improve--often substantially--detection performance in the experimental systems. 

\vspace{3pt}

\noindent \textbf{Overview of Experimental Setup.} We compare performance of privileged-augmented systems against two baseline (non-privileged) models: the standard set model and the complete set model. The \emph{standard set model} is a conventional detection system that does not include the privileged features for training or runtime, but uses all of the standard features. The \emph{complete set model} is a conventional system that includes all the privileged and standards features for training or runtime.  Note that the ideal privileged information approach would have similar performance as the complete set.

To learn standard and complete set models, we use classifiers of Random Forest  (RF) and Support Vector Machines (SVM) with a radial basis function kernel. These classifiers give better performance in the previously introduced systems and are also preferred by the system authors. The parameters of the models are optimized with exhaustive or randomized parameter search based on the dataset size. All of our experiments are implemented in Python with the scikit-learn machine learning library or MATLAB with the optimization toolbox and run on Intel i5 computer with 8 GB RAM. We give the details of the implementation of privileged-augmented systems while presenting the calibration of approaches in Section~\ref{sec:selectionApproach}. 

We show detection performance of complete and standard set models and compare their results with our approaches based on three metrics: accuracy, recall, and precision. We also present the false positives and negatives when relevant. \emph{Accuracy} is the sum of the true positive and true negatives over a total number of samples. \emph{Recall} is the number of true positives over the sum of false negatives and true positives, and \emph{precision} is the number of true positives over the sum of false positives and true positives. Higher values of accuracy, precision, and recall indicates a higher quality of the detection output. 

\begin{table}[t!]
\small
\centering
\resizebox{\columnwidth}{!}{%
\setlength{\tabcolsep}{6pt}
\renewcommand{\arraystretch}{0.5}
\begin{tabular}{@{}llccc@{}}
\cmidrule(l){3-5}
                                 &     & \multicolumn{3}{c}{\textbf{Fast-flux Bot Detection (FF)}} \\ \cmidrule(l){3-5} 
                                 &\textbf{Model}     & \textbf{Accuracy}          & \textbf{Precision}      & \textbf{Recall}     \\ \midrule
\multirow{2}{*}{Complete Set}    & RF  & 99$\pm$0.9      & 99.4           & 99.4       \\
                                 & SVM & 99.4$\pm$0.3       & 99.3          & 100        \\ \cmidrule(l){2-5} 
\multirow{2}{*}{Standard Set}    & RF  & 96.5$\pm$2.6       & 98.7          & 96.8      \\
                                 & SVM & 95$\pm$2.3       & 94.4          & 95.5       \\ \cmidrule(l){2-5} 
\multirow{2}{*}{Similarity (KT)} & RF  & \textbf{98.6$\pm$1.3}      & 99.3          & 98.7      \\
                                 & SVM & 96.7$\pm$1.2        & 98.7   & 97.1      \\ \cmidrule(l){2-5} 
\multirow{2}{*}{Regression (KT)} & RF  & 98.3$\pm$1.4      & 99          & 98.7      \\
                                 & SVM & 96$\pm$1.1       & 98.7         & 96.1      \\ \cmidrule(l){1-5} 
\end{tabular}
}
\caption{Fast-flux Bot Detection knowledge transfer (KT) results. All numbers shown are percentages. The best result of the knowledge transfer options is highlighted in boldface.}
\label{table:KTResults}
\end{table}

\subsection{Knowledge Transfer} 
\label{sec:evaluation-kt}   
Our first set of experiments compares the performance of privileged-augmented detection system using knowledge transfer (KT) over standard and complete set models.   In this experiment, we classify domain names into benign or malicious in fast-flux bot detection system (FF).

To realize KT, we implement two mapping functions to estimate the privileged features from a subset of standard features: \emph{regression-based} and \emph{similarity-based}. We find that both mapping functions learn the patterns of a training data and mostly suffice for the derivation of the nearly precise estimated privileged features. First, a polynomial regression function is built to find a coefficient vector $\beta \in \mathbb{R}^{d}$ such that there exists a $x_{i}^{*} = f_{i}({x}^{s}, \beta) + b + \epsilon$ for some bias term $b$ and random residual error $\epsilon$. The resulting function is then used to estimate each privileged feature at detection time given the standard features as an input. We use polynomial regression that fits a nonlinear relationship to each privileged feature and picks the one that minimizes the sum of squares error. To evaluate the effectiveness of the regression, we implement a second mapping function named weighted-similarity. This function is used to estimate the privileged features from the most similar samples in the training set. We first find the $k$ most similar subset of standard samples that are selected by using the Euclidean distance between an unknown sample and training instances. Then, the privileged features are replaced by assigning weights that are inversely proportional to the similarities of their neighbors. We note that other distance metrics gives worse accuracy than Euclidean distance for the studied datasets.

Table~\ref{table:KTResults} shows the accuracy of Random Forest Classifier (RF) and Support Vector Machines (SVM) on standard, complete models, and KT in the form of multiple regression and weighted-similarity. The average accuracy of ten independent runs of stratified cross-validation is given by measuring the difference between training and validation performance with parameter optimization (\eg $k$ parameter in similarity). The complete model accuracy of both classifiers is close to 99\%. Note that baseline performance is obtained by always guessing the most probable class yields 68\% accuracy. 

We found that mapping functions are effective in finding a nearly precise relation between standard and privileged features. This decreases the expected misclassification rate on average both in false positives and negatives over benchmark detection with no privileged features. Both KT mapping options come close to the complete model accuracy on FF dataset (1\% less accurate), and significantly exceeds the standard set accuracy (2\% more accurate). The results confirm that regression and similarity are more effective at estimating privileged features than solely using the standard features available at runtime.

\begin{table}[t!]
\small
\centering
\resizebox{\columnwidth}{!}{%
\setlength{\tabcolsep}{6pt}
\renewcommand{\arraystretch}{0.7}
\begin{tabular}{@{}llccc@{}}
\cmidrule(l){3-5}
\multicolumn{1}{l}{}          &                           & \multicolumn{3}{c}{\textbf{Malware Traffic Detection}}                                              \\ \cmidrule(l){3-5} 
\multicolumn{1}{l}{}          & \multicolumn{1}{c}{\textbf{Model}} & \multicolumn{1}{c}{\textbf{Accuracy}} & \multicolumn{1}{c}{\textbf{Precision}} & \multicolumn{1}{c}{\textbf{Recall}} \\\hline
\multirow{2}{*}{Complete Set} & RF                        & 98.7$\pm$0.3                & 99.7                         & 98.9                      \\
                              & SVM                       & 95.6$\pm$1.2                & 98.8 & 94.6                      \\ \cmidrule(l){2-5} 
\multirow{2}{*}{Standard Set} & RF                        & 92$\pm$3               & 97.4                          & 95.3                      \\
                              & SVM                       & 89.2$\pm$0.6              & 94                        & 94.6                      \\ \cmidrule(l){2-5} 
Model Influence               & SVM+                      & \textbf{94$\pm$1.4}       & 94.8                          & 98.8                      \\ \bottomrule
\end{tabular}}
\caption{Malware Traffic Detection model influence results. All numbers shown are percentages.}
\label{svmPlusTable}
\end{table}

\subsection{Model Influence}
\label{sec:ev-model-influence}
Next, we evaluate the performance of model influence-based privileged information in detecting Zeus botnet in real-world web (HTTP(S)) traffic. Here, the system attempts to detect the malicious activity of a Zeus botnet that connects to C\&C centers and filters private data. Note that the Zeus botnet uses HTTP mimicry to avoid detection. As a consequence, the sole use of standard features makes detection of Zeus difficult, resulting in high detection error (where Zeus traffic is mostly classified as legitimate web traffic). To this end, we include privileged features of packet flags, port numbers, and packet timing information from packet headers (See Appendix~\ref{sec:AppendixC} for the complete list of features).  We observe that while these features can be spoofed by adversaries under normal conditions, using them as privileged information may counteract spoofing (because inference does not consider their runtime value).

We evaluate accuracy gain of model influence over the standard model. We use a polynomial kernel in the objective function to perform a non-linear classification by implicitly mapping the features into a higher dimensional feature space. We note that we avoid overfitting by tuning of the regularization parameter. Table~\ref{svmPlusTable} presents the impact of model influence on accuracy, precision, and recall and compares with standard and complete models. We found that using the privileged features inherent to malicious and benign samples in model training systematically better separates the classes. This positive effect substantially improves both false negative and false positive rates.  The accuracy of model influence is close to the optimal accuracy and reduces the detection error on average 2\% over RF trained on the standard set. This positive effect is more observable in SVM, and the accuracy gain yields 4.8\%.

\subsection{Distillation}
\label{sec:ev-distillation}
We evaluate distillation on the experimental face authentication
system.  The standard features of a face authentication system consist of original images of users (\ie background included) with 3 RGB channels. We obtain the privileged features of each image from funneled and downscaled bounding boxes of images. In this way, we better characterize each user's face by specifying the face localization and eliminating the background noise. It is important to note that background of images may unrealistically increase the accuracy because background regions may contribute to the distinction between images. However, we verify by manual checking that the images in our training set do not suffer from this effect.
In the following experiments, we construct both standard and privileged models using a deep neural network (DNN) with two hidden layers of 10 rectifiers linear unit with a softmax layer for each class. This type of architecture is commonly applied in computer vision and provides superior results for image-specific applications. We train the network with ten runs of 400 random training samples.

\begin{figure}[t!]
\centering
\includegraphics[width=0.83\columnwidth]{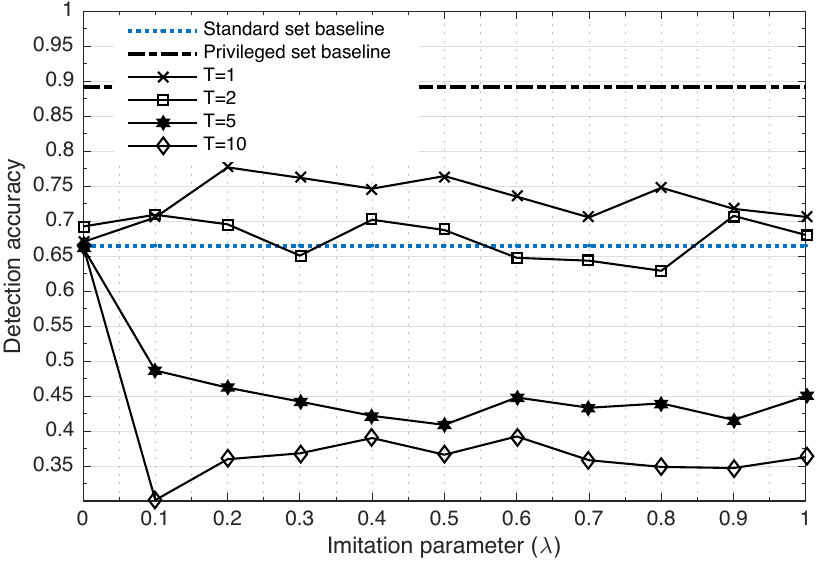}
\caption{Distillation impact on accuracy with privileged features in face authentication system. We plot standard and privileged set accuracy as a baseline. Temperature ($T$) and imitation ($\lambda$) parameters are quantified on various values to show their impact on accuracy.}
\label{fig:Gdistillation}
\end{figure}

Figure~\ref{fig:Gdistillation} plots the average distillation accuracy with various temperature and imitation parameters. We show the accuracy of standard  (dotted) and privileged set (dotted-dashed) models as a baseline. The resulting model achieves an average of 89.2\% correct classification rate on the privileged set, which is better than the standard set with 66.5\%. We observe that distilling the privileged set features into our detection algorithm gives better accuracy than standard set accuracy with optimal $T$ and $\lambda$ parameters. The accuracy is maximized when $T=1$, the gain is on average 6.56\%. The best improvement is obtained when $T=1$ and $\lambda=0.2$ with 11.2\% increase over the standard set model accuracy. However, increases in $T$ negatively affect detection accuracy.  This is because as $T$ increases, the objective function puts more weight on learning from the standard features which upsets the trade-off between standard and privileged features.

\begin{table*}[t!]
\small
\centering
\resizebox{\textwidth}{!}{%
\addtolength{\tabcolsep}{5pt}
\renewcommand{\arraystretch}{0.2}
\begin{tabular}{@{}lcccccccccc@{}}
\toprule
                & \multicolumn{1}{l}{} & \multicolumn{3}{c}{\textbf{Fast-flux Bot Detection}} & \multicolumn{3}{c}{\textbf{Malware Traffic Detection}} & \multicolumn{3}{c}{\textbf{Malware Classification}} \\ \midrule
                & \textbf{Model}                & \textbf{Acc.}               & \textbf{Pre.}      & \textbf{Rec.}      & \textbf{Acc.}                & \textbf{Pre.}       & \textbf{Rec.}       & \textbf{Acc.}              & \textbf{Pre.}      & \textbf{Rec.}     \\ \cmidrule(l){2-11} 
Complete Set    & RF                   & 99$\pm$0.9        & 99.4      & 99.4      & 98.7$\pm$0.3        & 99.7       & 98.9       & 96.6$\pm$1.2       & 99.3      & 95.2      \\ 
                & SVM                  & 99.4$\pm$0.3       & 99.3      & 100       & 95.6$\pm$1.2        & 98.8       & 94.6       & 95.7$\pm$1         & 98.6      & 94.6      \\ \cmidrule(l){2-11} 
Standard Set    & RF                   & 96.5$\pm$2.6       & 98.7      & 96.8      & 92.9$\pm$3          & 97.4       & 95.3       & 91.2$\pm$1         & 91.3      & 94.4      \\
                & SVM                  & 95$\pm$2.3       & 94.4      & 95.5      & 89.2$\pm$0.6        & 94         & 94.6       & 91.8$\pm$1.1       & 93.2      & 93.6      \\ \cmidrule(l){2-11} 
KT (Similarity) & RF                   & \bf{98.6$\pm$1.3}       & 99.3      & 98.7      & 93.3$\pm$2.1        & 94.4       & 98.4       & 90.1$\pm$2.2       & 97.3      & 88        \\
                & SVM                  & 96.7$\pm$1.2       & 98.7      & 97.1      & 92.6$\pm$0.9        & 95.1       & 96         & 83.5$\pm$11.2      & 88.3      & 85.6      \\ \cmidrule(l){2-11} 
Model Influence & SVM+                   & 97.3$\pm$1.3       & 97        & 99.3      & 94$\pm$1.4          & 94.8       & 98.8       &  \bf{94.6$\pm$2.3}      & 93.3      & 97.8      \\ \cmidrule(l){2-11} 
Distillation    & DNN                  & 97.5$\pm$0.3       & 97.4      & 99.3      &  \bf{95.7$\pm$0.6}        & 96.1       & 99.3       & 92.6$\pm$0.7       & 92.6      & 95.3      \\ \bottomrule
\end{tabular}}
\caption{Summary of results: Accuracy (Acc.), Precision (Pre.) and Recall (Rec.) The best result for each detection system is highlighted in bold.}
\label{table:all-results}
\end{table*}

\subsection{Comparison of Approaches}
\label{sec:model-comparison}
Next, we compare the relative performance of the approaches on three data sets.\footnote{We do not compare performance on face recognition because processing the number of the input features (e.g., pixels) was intractable for several solutions.}  Distillation is implemented using Deep Neural Networks (DNN), and regression and weighted similarity mapping functions are used for knowledge transfer. Table~\ref{table:all-results} presents the results of knowledge transfer (We report similarity results as it yields better results than regression), model influence and distillation and compares against complete and standard models.  The accuracy, precision and recall gain of the best resulting approach for each detection system is summarized in Table~\ref{summary-results}.

The accuracy of model influence, distillation and knowledge transfer on the fast-flux detector and malware traffic detection is stable. All approaches yield accuracy similar to the ideal accuracy of the complete set model, and often the increased accuracy is the result of the correct classification of true positives (intrusions). This results in on average up to 5\% relative gain in recall with the model influence and distillation over conventional models. In contrast, knowledge transfer often increases the number of samples detected by systems as being actually malicious (\eg 99.3\% precision in the fast-flux detector), meaning that the number of false alarms is reduced over conventional detection. The results confirm that the approaches are able to balance the conventional detection and its accuracy by using privileged features inherent to both benign and malicious samples: either reducing the false positives or negatives. We note that these results are obtained after carefully tuning the model parameters. We further discuss their parameter tuning in the following section and impact of results on systems in section~\ref{sec:discussion}.

Distillation is easy to apply as its objective is independent of the machine learning algorithm (model-free) and often yields better results than other approaches. Its quality as a detection mechanism becomes more apparent when its objective function is implemented with deep neural networks (DNN) with a nonlinear objective~\cite{lopez2015unifying}. This makes distillation give better results on average than other approaches. On the other hand, the design and calibration of model influence detection require additional effort and care in tuning parameters--in the experiments, this additional effort yields strong detection (as 94.6\% in malware classification). Note that when the dataset includes a large number of privileged features or samples, training of model influence takes significantly more time compared to other approaches (See next section).

Finally, it is important to note that the while knowledge transfer accuracy gain for fast-flux detection and malware traffic analysis is similar to other approaches, its malware classification results are inconsistent (\ie 83.5\% average accuracy with an 11.2\% standard deviation). Neither regression nor similarity mapping functions were able to predict the privileged features near precisely, in turn, they slightly degrade the accuracy (7-8\%) on both RF and SVM standard set models. This observation confirms the need to find and evaluate an appropriate mapping function for the transfer of knowledge discussed in Section~\ref{sec:knowledgeTransfer}. In this particular dataset, the mapping functions fail to find a good relation between standard and privileged features. Regression suffers from overfitting to uncommon data points and similarity lacks fitting data points that distinctly lie an abnormal distance from the range of standard features (confirmed by an increase in the sum of square errors of estimated and true values for the privileged features). We remark that derivation of more advanced mapping functions may solve this problem. Further, model influence and distillation solve this by eliminating the use of mapping functions and including standard and privileged feature dependency into their objectives. 

Therefore, based on the above observations, the approaches are in need for a calibration based on the domain and task-specific properties to maximize the detection gain, as explored next.

\section{Limitations} 
\label{sec:selectionApproach}
In this section, we discuss the required dataset properties, algorithm parameters, training, and runtime overhead of using privileged information for detection.  We also present guidelines and cautionary warnings for use of privileged information in realistic deployment environments. A summary of an approach selection criteria is presented in Table~\ref{table:approachSelection}.

\vspace{2pt}\noindent\textbf{Model Dependency.} Model selection is a task of picking an appropriate model (\eg classifier) to construct a detection function from a set of potential models. Knowledge transfer can be applied to a model of choice, as privileged features are inferred with any accurately selected mapping function. Distillation requires a model with a softmax output layer for obtaining probability vectors. However, we adapt model influence to SVM's objective function. 

\vspace{2pt}\noindent\textbf{Detection Overhead.} The mapping functions used in knowledge transfer may introduce detection delays while estimating the privileged features. For instance, weighted similarity introduced in Section~\ref{sec:knowledgeTransfer} defers estimation until detection without learning a function at training time (\ie lazy learner). This may introduce a detection bottleneck if dataset includes a large number of samples. To solve this problem, we apply stratified sampling to reduce the size of the dataset. Furthermore, constructing mapping functions at training time such as regression-based minimize the delay of estimating privileged features. For instance, in our experiments, weighted-similarity is used to estimate ten privileged features of 5K training samples in less than a second delay on 2.6GHz 2-core Intel i5 processor with 8GB RAM. Regression reduces this value to milliseconds. Therefore, if delay at runtime is the primary concern, we suggest using model influence and distillation for learning the detection model, as they introduce no overhead at runtime.

\begin{table}[t!]
\small
\begin{threeparttable}
\addtolength{\tabcolsep}{-3pt}
\renewcommand{\arraystretch}{0.8}
\resizebox{\columnwidth}{!}{
\centering
\begin{tabular}{@{}cccccc@{}}
\toprule
& \textbf{Model}      & \textbf{Detection time}         & \textbf{Mode}l          & \textbf{Training time} \\ 
\textbf{Approach}  & \textbf{dependency}    & \textbf{overhead}  & \textbf{optimization}    & \textbf{overhead}  \\
\cmidrule(rl){1-1}\cmidrule(rl){2-2}\cmidrule(rl){3-3}\cmidrule(rl){4-4}\cmidrule(l){5-5}
\multirow{1}{*}{Knowledge Transfer} & \xmark & \cmark & \pie{180}   & \pie{180} \\  
\multirow{1}{*}{Model Influence} & \cmark & \xmark & \pie{360}   & \pie{360}  \\  
\multirow{1}{*}{Distillation} & \xmark & \xmark &\pie{180} & \pie{180}    \\ 
\bottomrule
\end{tabular}
}
\label{table:approachComparision}
\vspace*{-1mm}
\begin{tablenotes}
  \item {\small \emph{Legend:} \cmark~yes \xmark~no $\pie{180}$ model dependent $\pie{360}$ relatively higher}
\end{tablenotes}
\end{threeparttable}
\caption{Guideline for approach selection.}
\label{table:approachSelection}
\end{table} 

\vspace{2pt}\noindent\textbf{Model Optimization.} To obtain the best performance results, the parameters and hyperparameters of approaches need to be carefully tuned. For instance, fine-tuning of temperature and imitation parameters in distillation and kernel function hyperparameters in model influence approaches may increase the detection performance. Similar to conventional detection, the number of parameters required to be optimized both for knowledge transfer and generalized distillation can be determined a priori based on the selected model. However, model influence has twice as many parameters as SVM---two kernel functions are used simultaneously to learn detection boundary in standard and privileged feature spaces. We apply grid search for small training sets and evolutionary search for large-scale datasets for parameter optimization. 

\vspace{2pt}\noindent\textbf{Training Overhead.} Training set size affects the time required by model learning. The amount of additional time needed to run both knowledge transfer and generalized distillation is negligible, as they require similar models as existing systems apply. However, the objective function of model influence may become infeasible or take a long time when the dimension of the feature space is very small, or dataset size is quite large. For instance, in our experiments, distillation and knowledge transfer train 1K samples with 50 standard and privileged features on the same machine used for one minute including optimal parameter search. Model influence takes on average 30 minutes on the same machine used for measuring detection overhead. Packages that are designed specifically for solving the quadratic programming (QP) problems (\eg MATLAB \texttt{quadprog()} function) can be used instead of general solvers such as convex optimization package CVX to reduce the training time. Further, specialized spline kernels can be used to accelerate the computation~\cite{vapnik2015learning}. We give a specific implementation of model influence in such packages in Appendix~\ref{sec:appendixA}.

\section{Discussion}
\label{sec:discussion}
Our empirical results show that approaches reduce both false positives and negatives over the systems solely built on their standard features. In a security setting, a false positive makes it extremely difficult for the analyst examining the reported incidents only to identify the mistakenly triggered benign events correctly. It is not surprising, therefore, that recent research focuses on post-processing of the alerts to produce a more qualitative alert set useful to the human analyst~\cite{shittu2015intrusion}. False negatives, on the other hand, have the potential to cause catastrophic damage to both users and organizations: even a single compromised system can cause serious security breaches. For instance, in malware traffic and fast-flux bot detection, a false negative may cause a bot filter private data to a malicious server.  In the case of malware classification and face authentication, it undermines the integrity of a system by misclassifying malware into another family or recognizing the wrong user. Thus, in no small way, improvement in false positives and negatives of these systems does matter in operational settings, improving reliable detection. 

\subsection{Uses of Privileged Information}
Privileged information is not restricted to the domains discussed above, but are readily adaptable problems and machine learning settings. For example, privileged information can be adapted to settings with unsupervised, regression, and metric learning~\cite{sideInformation}.  With respect to detection, we consider several illustrative uses of privileged information below:

\vspace{2pt}\noindent\textbf{Mobile Device Security} - The growth of mobile malware requires robust malware detectors on mobile devices. Current systems collect data for numerous type of attacks; however, exhaustive data collection at runtime can have high energy costs and induce noticeable interface lag.  As a consequence, users may disable the detection mechanism~\cite{bickford2011security}. 
We note that high-cost features can be defined as privileged information to combat this problem.  

\vspace{2pt}\noindent\textbf{Enterprise Security} - Enterprise systems use audit data generated from a diverse set of devices and information sources for analysis~\cite{cardenas2013big}. For instance, SIEM products 
collect data from hosts, applications, and network devices in incredible volumes (\eg 100K events per second yielding to 42 TB of compressed data). These massive datasets are mined for patterns identifying sophisticated threats and vulnerabilities.  However, systems may be overwhelmed by feature collection and processing at runtime which makes their collection impractical for many settings. In such cases, features involving complex and expensive data collection can be defined as privileged to balance the real-time costs and accuracy.

\vspace{2pt}\noindent\textbf{Privacy Enhanced Detection} - Many detection processes require the collection of privacy-relevant features, e.g., pattern and substance of user network traffic, use of the software~\cite{celik2016patient, curie}.  Hence, it is important to reduce the collection and exposure of such data--legal and ethical issues may prevent continuously monitoring them in their original form. In these cases, a set of features can be defined as privileged to eliminate the requirement of obtaining and potentially retaining privacy-sensitive features at runtime from users and environments.

\subsection{Privileged Information as a Defense}
\label{sec:adversarial}
We also posit that privileged information can be used as a defense mechanism in adversarial settings. More specifically, the key attacks targeting machine learning are organized into two categories based on adversarial capabilities~\cite{huang2011adversarial}: (\emph{1}) \emph{causative (poisoning) attacks} in which an attacker controls the training data by injecting well-crafted attack samples to control the prediction results, and (\emph{2}) \emph{explanatory (evasion) attacks} in which attacker  manipulates the malicious samples to evade detection. For the former, privileged features adds an extra step for the attacker to pollute the training data because the attacker needs to dupe the data collection into including polluted privileged samples in addition to the standard samples--which for many systems including online learning would potentially be much more difficult.   For the latter,  privileged features may make detection systems more robust to the adversarial samples because privileged features cannot be controlled by the adversary when producing malicious samples~\cite{laskov2014practical}.  Moreover, because the model is hidden from the adversary they cannot know the influence of these features on the model~\cite{biggio2014pattern}. As a proof of concept, recent works have used distillation of standard features as a defense mechanism against adversarial perturbations in DNNs~\cite{papernot2015distillation}.  In future work,  we plan to further evaluate privileged information as a mechanism to harden machine learning systems.

\section{Related Work}
Domain-specific feature engineering has been a key effort within the security communities.  For example, researchers have previously used specific patterns to group malware samples into families~\cite{rafique2013firma, nissim2014novel, ahmadi2015novel}, have explored using  DNS information to understand and predict botnet domains~\cite{antonakakis2012throw, bilge2011exposure, yadav2010detecting, celik2017malware}, and have analyzed network and system level features to identify previously unknown malware traffic~\cite{rahbarinia2015segugio,miller2012sequential,celik2011salting}. Other works have focused on user authentication from facial images~\cite{parkhi2015deep, sun2015deepid3}.  We view our efforts in this paper to be complementary to these and related works.  Features in these works can be easily enhanced with privileged information in detection algorithms to strike a balance between accuracy and the cost or availability constraints at runtime.
 
The use of privileged information has recently attracted attention in a few others areas such as computer vision, image processing, and even finance. Wang et al.~\cite{wang2015classifier} and Sharmanska et al.~\cite{sharmanska2013learning} derived privileged features from images in the form of annotator rationales, object bounding boxes, and textual descriptions. 
Ribeiro et al. used annual turnover and global balance values as privileged features for enhancing the financial decision-making~\cite{ribeiro2012enhanced}. However, their approaches are not designed to model security-relevant data and do not consider feature engineering but rather to determine if there is a possibility of application to a domain specific information.

\section{Conclusions}
We have presented a range of techniques to train detection systems with privileged information.  All approaches use features available only at training time to enhance the accuracy of detection models.  We consider three approaches: (a) knowledge transfer to construct mapping functions to estimate the privileged features, (b)  model influence to smooth the detection model with the useful information obtained from the privileged features, and (c) distillation using probability vector outputs obtained from the privileged features in the detection objective function. Our evaluation of several detection systems shows that we can we improve the accuracy, recall, and precision regardless of their high detection performance using privileged features. We also presented guidelines for approach selection in realistic deployment environments. 

This work is the first effort at developing detection under privileged information by exploring feature engineering, algorithms, and environmental calibration. The capability afforded by this approach will allow us to integrate forensic and other auxiliary information that, to date, has not been actionable for detection.  In the future, we will explore various environments and evaluate its ability to promote resilience to adversarial manipulation in detection systems.  In this way, we will explore new models and systems using privileged features to promote lightweight, accurate and robust detection.

\begin{acks}
Research was sponsored by the Army Research Laboratory and was accomplished under Cooperative Agreement Number W911NF-13-2-0045 (ARL Cyber Security CRA). The views and conclusions contained in this document are those of the authors and should not be interpreted as representing the official policies, either expressed or implied, of the Army Research Laboratory or the U.S. Government. The U.S. Government is authorized to reproduce and distribute reprints for Government purposes notwithstanding any copyright notation here on.
\end{acks}

\bibliographystyle{ACM-Reference-Format}
\bibliography{arxiv-asiaaccs-extended-privileged.bbl}


\begin{thebibliography}{60}


\ifx \showCODEN    \undefined \def \showCODEN     #1{\unskip}     \fi
\ifx \showDOI      \undefined \def \showDOI       #1{#1}\fi
\ifx \showISBNx    \undefined \def \showISBNx     #1{\unskip}     \fi
\ifx \showISBNxiii \undefined \def \showISBNxiii  #1{\unskip}     \fi
\ifx \showISSN     \undefined \def \showISSN      #1{\unskip}     \fi
\ifx \showLCCN     \undefined \def \showLCCN      #1{\unskip}     \fi
\ifx \shownote     \undefined \def \shownote      #1{#1}          \fi
\ifx \showarticletitle \undefined \def \showarticletitle #1{#1}   \fi
\ifx \showURL      \undefined \def \showURL       {\relax}        \fi
\providecommand\bibfield[2]{#2}
\providecommand\bibinfo[2]{#2}
\providecommand\natexlab[1]{#1}
\providecommand\showeprint[2][]{arXiv:#2}

\bibitem[\protect\citeauthoryear{A. et~al\mbox{.}}{A. et~al\mbox{.}}{2012}]%
        {antonakakis2012throw}
\bibfield{author}{\bibinfo{person}{Manos A.} {et~al\mbox{.}}}
  \bibinfo{year}{2012}\natexlab{}.
\newblock \showarticletitle{From throw-away traffic to bots: detecting the rise
  of DGA-based malware}. In \bibinfo{booktitle}{\emph{USENIX Security}}.
\newblock


\bibitem[\protect\citeauthoryear{A. et~al\mbox{.}}{A. et~al\mbox{.}}{2015}]%
        {ahmadi2015novel}
\bibfield{author}{\bibinfo{person}{Mansour A.} {et~al\mbox{.}}}
  \bibinfo{year}{2015}\natexlab{}.
\newblock \showarticletitle{Novel feature extraction, selection and fusion for
  effective malware family classification}.
\newblock \bibinfo{journal}{\emph{arXiv preprint arXiv:1511.04317}}.
\newblock


\bibitem[\protect\citeauthoryear{B. and C.}{B. and C.}{2014}]%
        {ba2014deep}
\bibfield{author}{\bibinfo{person}{Jimmy B.} {and} \bibinfo{person}{Rich C.}}
  \bibinfo{year}{2014}\natexlab{}.
\newblock \showarticletitle{Do deep nets really need to be deep?}. In
  \bibinfo{booktitle}{\emph{Advances in neural information processing
  systems}}.
\newblock


\bibitem[\protect\citeauthoryear{B., Kirda, Kruegel, and B.}{B.
  et~al\mbox{.}}{2011}]%
        {bilge2011exposure}
\bibfield{author}{\bibinfo{person}{Leyla B.}, \bibinfo{person}{E. Kirda},
  \bibinfo{person}{C. Kruegel}, {and} \bibinfo{person}{Marco B.}}
  \bibinfo{year}{2011}\natexlab{}.
\newblock \showarticletitle{EXPOSURE: Finding malicious domains using passive
  {DNS} analysis}. In \bibinfo{booktitle}{\emph{NDSS}}.
\newblock


\bibitem[\protect\citeauthoryear{Barbosa, Sadre, Pras, and Meent}{Barbosa
  et~al\mbox{.}}{2010}]%
        {barbosa2010simpleweb}
\bibfield{author}{\bibinfo{person}{R. Barbosa}, \bibinfo{person}{R. Sadre},
  \bibinfo{person}{A. Pras}, {and} \bibinfo{person}{R. Meent}.}
  \bibinfo{year}{2010}\natexlab{}.
\newblock \showarticletitle{University of Ttwente traffic traces data
  repository}. In \bibinfo{booktitle}{\emph{University of Twente Tech Report}}.
\newblock


\bibitem[\protect\citeauthoryear{Bickford et~al\mbox{.}}{Bickford
  et~al\mbox{.}}{2011}]%
        {bickford2011security}
\bibfield{author}{\bibinfo{person}{J. Bickford} {et~al\mbox{.}}}
  \bibinfo{year}{2011}\natexlab{}.
\newblock \showarticletitle{Security versus energy tradeoffs in host-based
  mobile malware detection}. In \bibinfo{booktitle}{\emph{Mobile systems,
  applications, and services}}.
\newblock


\bibitem[\protect\citeauthoryear{Biggio, Fumera, and Roli}{Biggio
  et~al\mbox{.}}{2014}]%
        {biggio2014pattern}
\bibfield{author}{\bibinfo{person}{B. Biggio}, \bibinfo{person}{G. Fumera},
  {and} \bibinfo{person}{F. Roli}.} \bibinfo{year}{2014}\natexlab{}.
\newblock \showarticletitle{Pattern recognition systems under attack: Design
  issues and research challenges}.
\newblock \bibinfo{journal}{\emph{International Journal of Pattern Recognition
  and Artificial Intelligence}}.
\newblock


\bibitem[\protect\citeauthoryear{Breach}{Breach}{2018}]%
        {Equifax}
\bibfield{author}{\bibinfo{person}{Equifax~Data Breach}.}
  \bibinfo{year}{2018}\natexlab{}.
\newblock \bibinfo{howpublished}{{https://en.wikipedia.org/wiki/Equifax}}.
  (\bibinfo{year}{2018}).
\newblock
\newblock
\shownote{[Online; accessed 15-January-2018].}


\bibitem[\protect\citeauthoryear{Butun et~al\mbox{.}}{Butun
  et~al\mbox{.}}{2014}]%
        {butun2014survey}
\bibfield{author}{\bibinfo{person}{I. Butun} {et~al\mbox{.}}}
  \bibinfo{year}{2014}\natexlab{}.
\newblock \showarticletitle{A survey of intrusion detection systems in wireless
  sensor networks}.
\newblock \bibinfo{journal}{\emph{IEEE Communications Surveys \& Tutorials}}.
\newblock


\bibitem[\protect\citeauthoryear{Cardenas, Manadhata, and Rajan}{Cardenas
  et~al\mbox{.}}{2013}]%
        {cardenas2013big}
\bibfield{author}{\bibinfo{person}{A.~A. Cardenas}, \bibinfo{person}{P.~K.
  Manadhata}, {and} \bibinfo{person}{S.~P. Rajan}.}
  \bibinfo{year}{2013}\natexlab{}.
\newblock \showarticletitle{Big data analytics for security}.
\newblock \bibinfo{journal}{\emph{IEEE System Security}}.
\newblock


\bibitem[\protect\citeauthoryear{Celik, Aksu, Acar, Sheatsley, Uluagac, and
  McDaniel}{Celik et~al\mbox{.}}{2017a}]%
        {curie}
\bibfield{author}{\bibinfo{person}{Z.~B. Celik}, \bibinfo{person}{H. Aksu},
  \bibinfo{person}{A. Acar}, \bibinfo{person}{R. Sheatsley},
  \bibinfo{person}{A.~S. Uluagac}, {and} \bibinfo{person}{P. McDaniel}.}
  \bibinfo{year}{2017}\natexlab{a}.
\newblock \showarticletitle{Curie: Policy-based Secure Data Exchange}.
\newblock
\showeprint{arXiv:1702.08342}


\bibitem[\protect\citeauthoryear{Celik, Izmailov, and McDaniel}{Celik
  et~al\mbox{.}}{2015a}]%
        {techReportLupi}
\bibfield{author}{\bibinfo{person}{Z.~B. Celik}, \bibinfo{person}{R. Izmailov},
  {and} \bibinfo{person}{P. McDaniel}.} \bibinfo{year}{2015}\natexlab{a}.
\newblock \bibinfo{booktitle}{\emph{{Proof and implementation of algorithmic
  realization of learning using privileged information (LUPI) paradigm:
  SVM+}}}.
\newblock \bibinfo{type}{{T}echnical {R}eport} NAS-TR-0187-2015.
  \bibinfo{institution}{{NSCR}}, \bibinfo{address}{CSE, PSU}.
\newblock


\bibitem[\protect\citeauthoryear{Celik, Lopez-Paz, and McDaniel}{Celik
  et~al\mbox{.}}{2016}]%
        {celik2016patient}
\bibfield{author}{\bibinfo{person}{Z.~Berkay Celik}, \bibinfo{person}{David
  Lopez-Paz}, {and} \bibinfo{person}{Patrick McDaniel}.}
  \bibinfo{year}{2016}\natexlab{}.
\newblock \showarticletitle{Patient-driven privacy control through generalized
  distillation}.
\newblock \bibinfo{journal}{\emph{IEEE Symposium on Privacy-Aware Computing
  (PAC)}}.
\newblock


\bibitem[\protect\citeauthoryear{Celik, McDaniel, and Bowen}{Celik
  et~al\mbox{.}}{2017b}]%
        {celik2017malware}
\bibfield{author}{\bibinfo{person}{Z.~B. Celik}, \bibinfo{person}{P. McDaniel},
  {and} \bibinfo{person}{T. Bowen}.} \bibinfo{year}{2017}\natexlab{b}.
\newblock \showarticletitle{Malware modeling and experimentation through
  parameterized behavior}.
\newblock \bibinfo{journal}{\emph{The Journal of Defense Modeling and
  Simulation}}.
\newblock


\bibitem[\protect\citeauthoryear{Celik, McDaniel, and Izmailov}{Celik
  et~al\mbox{.}}{2017c}]%
        {celik2017feature}
\bibfield{author}{\bibinfo{person}{Z.~B. Celik}, \bibinfo{person}{P. McDaniel},
  {and} \bibinfo{person}{R. Izmailov}.} \bibinfo{year}{2017}\natexlab{c}.
\newblock \showarticletitle{Feature cultivation in privileged
  information-augmented detection}. In \bibinfo{booktitle}{\emph{ACM CODASPY
  International Workshop on Security And Privacy Analytics}}.
\newblock


\bibitem[\protect\citeauthoryear{Celik and Oktug}{Celik and Oktug}{2013}]%
        {celik2013detection}
\bibfield{author}{\bibinfo{person}{Z.~B. Celik} {and} \bibinfo{person}{S.
  Oktug}.} \bibinfo{year}{2013}\natexlab{}.
\newblock \showarticletitle{{Detection of fast-flux networks using various DNS
  feature sets}}. In \bibinfo{booktitle}{\emph{ISCC}}.
\newblock


\bibitem[\protect\citeauthoryear{Celik, Raghuram, Kesidis, and Miller}{Celik
  et~al\mbox{.}}{2011}]%
        {celik2011salting}
\bibfield{author}{\bibinfo{person}{Z.~B. Celik}, \bibinfo{person}{J. Raghuram},
  \bibinfo{person}{G. Kesidis}, {and} \bibinfo{person}{D.~J. Miller}.}
  \bibinfo{year}{2011}\natexlab{}.
\newblock \showarticletitle{Salting public traces with attack traffic to test
  flow classifiers}. In \bibinfo{booktitle}{\emph{Usenix CSET}}.
\newblock


\bibitem[\protect\citeauthoryear{Celik, Walls, McDaniel, and Swami}{Celik
  et~al\mbox{.}}{2015b}]%
        {berkay2015malware}
\bibfield{author}{\bibinfo{person}{Z.~B. Celik}, \bibinfo{person}{R.~J. Walls},
  \bibinfo{person}{P. McDaniel}, {and} \bibinfo{person}{A. Swami}.}
  \bibinfo{year}{2015}\natexlab{b}.
\newblock \showarticletitle{Malware traffic detection using tamper resistant
  features}. In \bibinfo{booktitle}{\emph{IEEE MILCOM}}.
\newblock


\bibitem[\protect\citeauthoryear{Challenge}{Challenge}{2017}]%
        {kaggleMicrosoft}
\bibfield{author}{\bibinfo{person}{Microsoft Malware~Classification
  Challenge}.} \bibinfo{year}{2017}\natexlab{}.
\newblock
  \bibinfo{howpublished}{https://www.kaggle.com/c/malware-classification/}.
  (\bibinfo{year}{2017}).
\newblock
\newblock
\shownote{[Online; accessed 10-May-2017].}


\bibitem[\protect\citeauthoryear{Duc and Minh}{Duc and Minh}{2009}]%
        {duc2009your}
\bibfield{author}{\bibinfo{person}{N.~M. Duc} {and} \bibinfo{person}{B.~Q.
  Minh}.} \bibinfo{year}{2009}\natexlab{}.
\newblock \showarticletitle{Your face is not your password face authentication
  bypassing lenovo--asus--toshiba}.
\newblock \bibinfo{journal}{\emph{Black Hat Briefings}}.
\newblock


\bibitem[\protect\citeauthoryear{Garc{\'\i}a, Grill, Stiborek, and
  Zunino}{Garc{\'\i}a et~al\mbox{.}}{2014}]%
        {garcia2014empirical}
\bibfield{author}{\bibinfo{person}{S. Garc{\'\i}a}, \bibinfo{person}{M. Grill},
  \bibinfo{person}{J. Stiborek}, {and} \bibinfo{person}{A. Zunino}.}
  \bibinfo{year}{2014}\natexlab{}.
\newblock \showarticletitle{An empirical comparison of botnet detection
  methods}.
\newblock \bibinfo{journal}{\emph{In Computers \& Security}}.
\newblock


\bibitem[\protect\citeauthoryear{Gu et~al\mbox{.}}{Gu et~al\mbox{.}}{2008}]%
        {gu2008botminer}
\bibfield{author}{\bibinfo{person}{G. Gu} {et~al\mbox{.}}}
  \bibinfo{year}{2008}\natexlab{}.
\newblock \showarticletitle{BotMiner: Clustering Analysis of Network Traffic
  for Protocol-and Structure-Independent Botnet Detection}. In
  \bibinfo{booktitle}{\emph{USENIX Security}}.
\newblock


\bibitem[\protect\citeauthoryear{Hassanzadeh et~al\mbox{.}}{Hassanzadeh
  et~al\mbox{.}}{2013}]%
        {hassanzadeh2013pride}
\bibfield{author}{\bibinfo{person}{A. Hassanzadeh} {et~al\mbox{.}}}
  \bibinfo{year}{2013}\natexlab{}.
\newblock \showarticletitle{PRIDE: Practical intrusion detection in resource
  constrained wireless mesh networks}. In \bibinfo{booktitle}{\emph{Information
  and Communications Security}}.
\newblock


\bibitem[\protect\citeauthoryear{Hinton, Vinyals, and Dean}{Hinton
  et~al\mbox{.}}{2015}]%
        {hinton2015distilling}
\bibfield{author}{\bibinfo{person}{G. Hinton}, \bibinfo{person}{O. Vinyals},
  {and} \bibinfo{person}{J. Dean}.} \bibinfo{year}{2015}\natexlab{}.
\newblock \showarticletitle{Distilling the knowledge in a neural network}.
\newblock \bibinfo{journal}{\emph{arXiv preprint arXiv:1503.02531}}.
\newblock


\bibitem[\protect\citeauthoryear{Hsu, Huang, et~al\mbox{.}}{Hsu
  et~al\mbox{.}}{2010}]%
        {hsu2010fast}
\bibfield{author}{\bibinfo{person}{C. Hsu}, \bibinfo{person}{C. Huang},
  {et~al\mbox{.}}} \bibinfo{year}{2010}\natexlab{}.
\newblock \showarticletitle{Fast-flux bot detection in real time}. In
  \bibinfo{booktitle}{\emph{RAID}}.
\newblock


\bibitem[\protect\citeauthoryear{Huang et~al\mbox{.}}{Huang
  et~al\mbox{.}}{2012}]%
        {huang2012learning}
\bibfield{author}{\bibinfo{person}{Gary Huang} {et~al\mbox{.}}}
  \bibinfo{year}{2012}\natexlab{}.
\newblock \showarticletitle{Learning to align from scratch}. In
  \bibinfo{booktitle}{\emph{Advances in Neural Information Processing
  Systems}}.
\newblock


\bibitem[\protect\citeauthoryear{Huang et~al\mbox{.}}{Huang
  et~al\mbox{.}}{2014}]%
        {LFWTechUpdate}
\bibfield{author}{\bibinfo{person}{G.~B. Huang} {et~al\mbox{.}}}
  \bibinfo{year}{2014}\natexlab{}.
\newblock \bibinfo{booktitle}{\emph{Labeled faces in the wild: Updates and new
  reporting procedures}}.
\newblock \bibinfo{type}{{T}echnical {R}eport} UM-CS-2014-003.
  \bibinfo{institution}{UMASS}.
\newblock


\bibitem[\protect\citeauthoryear{Huang et~al\mbox{.}}{Huang
  et~al\mbox{.}}{2011}]%
        {huang2011adversarial}
\bibfield{author}{\bibinfo{person}{L. Huang} {et~al\mbox{.}}}
  \bibinfo{year}{2011}\natexlab{}.
\newblock \showarticletitle{Adversarial machine learning}. In
  \bibinfo{booktitle}{\emph{ACM security and artificial intelligence
  workshop}}.
\newblock


\bibitem[\protect\citeauthoryear{Huang et~al\mbox{.}}{Huang
  et~al\mbox{.}}{2010}]%
        {huang2010fast}
\bibfield{author}{\bibinfo{person}{S. Huang} {et~al\mbox{.}}}
  \bibinfo{year}{2010}\natexlab{}.
\newblock \showarticletitle{Fast-flux service network detection based on
  spatial snapshot mechanism for delay-free detection}. In
  \bibinfo{booktitle}{\emph{ASIACCS}}.
\newblock


\bibitem[\protect\citeauthoryear{J. and Learned-Miller}{J. and
  Learned-Miller}{2010}]%
        {fddbTech}
\bibfield{author}{\bibinfo{person}{Vidit J.} {and} \bibinfo{person}{E.
  Learned-Miller}.} \bibinfo{year}{2010}\natexlab{}.
\newblock \bibinfo{booktitle}{\emph{FDDB: A Benchmark for Face Detection in
  Unconstrained Settings}}.
\newblock \bibinfo{type}{{T}echnical {R}eport} UM-CS-2010-009.
  \bibinfo{institution}{UMass}.
\newblock


\bibitem[\protect\citeauthoryear{Jonschkowski, Höfer, and Brock}{Jonschkowski
  et~al\mbox{.}}{2015}]%
        {sideInformation}
\bibfield{author}{\bibinfo{person}{Rico Jonschkowski},
  \bibinfo{person}{Sebastian Höfer}, {and} \bibinfo{person}{Oliver Brock}.}
  \bibinfo{year}{2015}\natexlab{}.
\newblock \showarticletitle{Patterns for learning with side information}.
\newblock
\showeprint{arXiv:1511.06429}


\bibitem[\protect\citeauthoryear{Laskov et~al\mbox{.}}{Laskov
  et~al\mbox{.}}{2014}]%
        {laskov2014practical}
\bibfield{author}{\bibinfo{person}{P. Laskov} {et~al\mbox{.}}}
  \bibinfo{year}{2014}\natexlab{}.
\newblock \showarticletitle{Practical evasion of a learning-based classifier: A
  case study}. In \bibinfo{booktitle}{\emph{IEEE Security and Privacy}}.
\newblock


\bibitem[\protect\citeauthoryear{Lopez-Paz, Bottou, Sch{\"o}lkopf, and
  Vapnik}{Lopez-Paz et~al\mbox{.}}{2015}]%
        {lopez2015unifying}
\bibfield{author}{\bibinfo{person}{D. Lopez-Paz}, \bibinfo{person}{L. Bottou},
  \bibinfo{person}{B. Sch{\"o}lkopf}, {and} \bibinfo{person}{V. Vapnik}.}
  \bibinfo{year}{2015}\natexlab{}.
\newblock \showarticletitle{Unifying distillation and privileged information}.
\newblock \bibinfo{journal}{\emph{arXiv preprint arXiv:1511.03643}}.
\newblock


\bibitem[\protect\citeauthoryear{{MATLAB documentation of quadprog
  function}}{{MATLAB documentation of quadprog function}}{2018}]%
        {matlabQuadprog}
\bibfield{author}{\bibinfo{person}{{MATLAB documentation of quadprog
  function}}.} \bibinfo{year}{2018}\natexlab{}.
\newblock
  \bibinfo{howpublished}{\url{www.mathworks.com/help/optim/ug/quadprog.html}}.
\newblock
\newblock
\shownote{[Online; accessed 15-March-2018].}


\bibitem[\protect\citeauthoryear{McDaniel, Papernot, and Celik}{McDaniel
  et~al\mbox{.}}{2016}]%
        {mcdaniel2016machine}
\bibfield{author}{\bibinfo{person}{P. McDaniel}, \bibinfo{person}{N. Papernot},
  {and} \bibinfo{person}{Z.~B. Celik}.} \bibinfo{year}{2016}\natexlab{}.
\newblock \showarticletitle{Machine learning in adversarial settings}.
\newblock \bibinfo{journal}{\emph{Security \& Privacy Magazine}}.
\newblock


\bibitem[\protect\citeauthoryear{Miller et~al\mbox{.}}{Miller
  et~al\mbox{.}}{2012}]%
        {miller2012sequential}
\bibfield{author}{\bibinfo{person}{D.~J. Miller} {et~al\mbox{.}}}
  \bibinfo{year}{2012}\natexlab{}.
\newblock \showarticletitle{Sequential anomaly detection in a batch with
  growing number of tests: Application to network intrusion detection}. In
  \bibinfo{booktitle}{\emph{MLSP}}.
\newblock


\bibitem[\protect\citeauthoryear{Nataraj, Yegneswaran, Porras, and
  Zhang}{Nataraj et~al\mbox{.}}{2011}]%
        {nataraj2011comparative}
\bibfield{author}{\bibinfo{person}{L. Nataraj}, \bibinfo{person}{V.
  Yegneswaran}, \bibinfo{person}{P. Porras}, {and} \bibinfo{person}{J. Zhang}.}
  \bibinfo{year}{2011}\natexlab{}.
\newblock \showarticletitle{A comparative assessment of malware classification
  using binary texture analysis and dynamic analysis}. In
  \bibinfo{booktitle}{\emph{Security and Artificial Intelligence Workshop}}.
  ACM.
\newblock


\bibitem[\protect\citeauthoryear{Nissim et~al\mbox{.}}{Nissim
  et~al\mbox{.}}{2014}]%
        {nissim2014novel}
\bibfield{author}{\bibinfo{person}{N. Nissim} {et~al\mbox{.}}}
  \bibinfo{year}{2014}\natexlab{}.
\newblock \showarticletitle{Novel active learning methods for enhanced PC
  malware detection in windows OS}.
\newblock \bibinfo{journal}{\emph{Expert Systems with Applications}}.
\newblock


\bibitem[\protect\citeauthoryear{Pang, Allman, Bennett, Lee, Paxson, and
  Tierney}{Pang et~al\mbox{.}}{2005}]%
        {pang2005first}
\bibfield{author}{\bibinfo{person}{R. Pang}, \bibinfo{person}{M. Allman},
  \bibinfo{person}{M. Bennett}, \bibinfo{person}{J. Lee}, \bibinfo{person}{V.
  Paxson}, {and} \bibinfo{person}{B. Tierney}.}
  \bibinfo{year}{2005}\natexlab{}.
\newblock \showarticletitle{A first look at modern enterprise traffic}. In
  \bibinfo{booktitle}{\emph{Internet Measurement Conference}}.
\newblock


\bibitem[\protect\citeauthoryear{Papernot et~al\mbox{.}}{Papernot
  et~al\mbox{.}}{2016}]%
        {papernot2015distillation}
\bibfield{author}{\bibinfo{person}{N. Papernot} {et~al\mbox{.}}}
  \bibinfo{year}{2016}\natexlab{}.
\newblock \showarticletitle{Distillation as a defense to adversarial
  perturbations against deep neural networks}.
\newblock \bibinfo{journal}{\emph{IEEE S\&P}}.
\newblock


\bibitem[\protect\citeauthoryear{Parkhi et~al\mbox{.}}{Parkhi
  et~al\mbox{.}}{2015}]%
        {parkhi2015deep}
\bibfield{author}{\bibinfo{person}{O.~M. Parkhi} {et~al\mbox{.}}}
  \bibinfo{year}{2015}\natexlab{}.
\newblock \showarticletitle{Deep face recognition}.
\newblock \bibinfo{journal}{\emph{British Machine Vision}}.
\newblock


\bibitem[\protect\citeauthoryear{Passerini, Paleari, Martignoni, and
  Bruschi}{Passerini et~al\mbox{.}}{2008}]%
        {passerini2008fluxor}
\bibfield{author}{\bibinfo{person}{E. Passerini}, \bibinfo{person}{R. Paleari},
  \bibinfo{person}{L. Martignoni}, {and} \bibinfo{person}{D. Bruschi}.}
  \bibinfo{year}{2008}\natexlab{}.
\newblock \showarticletitle{Fluxor: Detecting and monitoring fast-flux service
  networks}. In \bibinfo{booktitle}{\emph{DIMVA}}.
\newblock


\bibitem[\protect\citeauthoryear{Rafique and Caballero}{Rafique and
  Caballero}{2013}]%
        {rafique2013firma}
\bibfield{author}{\bibinfo{person}{M.~Z. Rafique} {and} \bibinfo{person}{J.
  Caballero}.} \bibinfo{year}{2013}\natexlab{}.
\newblock \showarticletitle{Firma: Malware clustering and network signature
  generation with mixed network behaviors}. In
  \bibinfo{booktitle}{\emph{RAID}}.
\newblock


\bibitem[\protect\citeauthoryear{Rahbarinia et~al\mbox{.}}{Rahbarinia
  et~al\mbox{.}}{2015}]%
        {rahbarinia2015segugio}
\bibfield{author}{\bibinfo{person}{B. Rahbarinia} {et~al\mbox{.}}}
  \bibinfo{year}{2015}\natexlab{}.
\newblock \showarticletitle{Segugio: Efficient behavior-based tracking of
  malware-control domains in large ISP networks}. In
  \bibinfo{booktitle}{\emph{IEEE DSN}}.
\newblock


\bibitem[\protect\citeauthoryear{Raza, Wallgren, and Voigt}{Raza
  et~al\mbox{.}}{2013}]%
        {raza2013svelte}
\bibfield{author}{\bibinfo{person}{S. Raza}, \bibinfo{person}{L. Wallgren},
  {and} \bibinfo{person}{T. Voigt}.} \bibinfo{year}{2013}\natexlab{}.
\newblock \showarticletitle{SVELTE: Real-time intrusion detection in the
  Internet of Things}.
\newblock \bibinfo{journal}{\emph{Ad hoc networks}}.
\newblock


\bibitem[\protect\citeauthoryear{Reeves, Ramaswamy, Locasto, Bratus, and
  Smith}{Reeves et~al\mbox{.}}{2011}]%
        {reeves2011lightweight}
\bibfield{author}{\bibinfo{person}{J. Reeves}, \bibinfo{person}{A. Ramaswamy},
  \bibinfo{person}{M. Locasto}, \bibinfo{person}{S. Bratus}, {and}
  \bibinfo{person}{S. Smith}.} \bibinfo{year}{2011}\natexlab{}.
\newblock \showarticletitle{Lightweight intrusion detection for
  resource-constrained embedded control systems}. In
  \bibinfo{booktitle}{\emph{Critical Infrastructure Protection}}.
\newblock


\bibitem[\protect\citeauthoryear{Ribeiro et~al\mbox{.}}{Ribeiro
  et~al\mbox{.}}{2012}]%
        {ribeiro2012enhanced}
\bibfield{author}{\bibinfo{person}{Bernardete Ribeiro} {et~al\mbox{.}}}
  \bibinfo{year}{2012}\natexlab{}.
\newblock \showarticletitle{Enhanced default risk models with SVM+}.
\newblock \bibinfo{journal}{\emph{Expert Systems with Applications}}.
\newblock


\bibitem[\protect\citeauthoryear{Rossow et~al\mbox{.}}{Rossow
  et~al\mbox{.}}{2013}]%
        {rossow2013sok}
\bibfield{author}{\bibinfo{person}{C. Rossow} {et~al\mbox{.}}}
  \bibinfo{year}{2013}\natexlab{}.
\newblock \showarticletitle{Sok: P2pwned-modeling and evaluating the resilience
  of peer-to-peer botnets}. In \bibinfo{booktitle}{\emph{IEEE Security and
  Privacy}}.
\newblock


\bibitem[\protect\citeauthoryear{Scarfone and Mell}{Scarfone and Mell}{2007}]%
        {scarfone2007guide}
\bibfield{author}{\bibinfo{person}{K. Scarfone} {and} \bibinfo{person}{P.
  Mell}.} \bibinfo{year}{2007}\natexlab{}.
\newblock \showarticletitle{Guide to intrusion detection and prevention systems
  ({IDPS})}.
\newblock \bibinfo{journal}{\emph{NIST special publication}}.
\newblock


\bibitem[\protect\citeauthoryear{Sharmanska, Quadrianto, and
  Lampert}{Sharmanska et~al\mbox{.}}{2013}]%
        {sharmanska2013learning}
\bibfield{author}{\bibinfo{person}{V. Sharmanska}, \bibinfo{person}{N.
  Quadrianto}, {and} \bibinfo{person}{C.~H. Lampert}.}
  \bibinfo{year}{2013}\natexlab{}.
\newblock \showarticletitle{Learning to rank using privileged information}. In
  \bibinfo{booktitle}{\emph{International Conference on Computer Vision}}.
\newblock


\bibitem[\protect\citeauthoryear{Shittu et~al\mbox{.}}{Shittu
  et~al\mbox{.}}{2015}]%
        {shittu2015intrusion}
\bibfield{author}{\bibinfo{person}{R. Shittu} {et~al\mbox{.}}}
  \bibinfo{year}{2015}\natexlab{}.
\newblock \showarticletitle{Intrusion alert prioritisation and attack detection
  using post-correlation analysis}.
\newblock \bibinfo{journal}{\emph{Computers \& Security}}.
\newblock


\bibitem[\protect\citeauthoryear{Sun, Liang, Wang, and Tang}{Sun
  et~al\mbox{.}}{2015}]%
        {sun2015deepid3}
\bibfield{author}{\bibinfo{person}{Y. Sun}, \bibinfo{person}{D. Liang},
  \bibinfo{person}{X. Wang}, {and} \bibinfo{person}{X. Tang}.}
  \bibinfo{year}{2015}\natexlab{}.
\newblock \showarticletitle{Deepid3: Face recognition with very DNNs}.
\newblock \bibinfo{journal}{\emph{arXiv preprint arXiv:1502.00873}}.
\newblock


\bibitem[\protect\citeauthoryear{Turlach and Weingessel}{Turlach and
  Weingessel}{2007}]%
        {turlach2007quadprog}
\bibfield{author}{\bibinfo{person}{B.~A. Turlach} {and} \bibinfo{person}{A
  Weingessel}.} \bibinfo{year}{2007}\natexlab{}.
\newblock \showarticletitle{quadprog: Functions to solve quadratic programming
  problems}.
\newblock \bibinfo{journal}{\emph{R package version}}.
\newblock


\bibitem[\protect\citeauthoryear{Vapnik and Izmailov}{Vapnik and
  Izmailov}{2015}]%
        {vapnik2015learning}
\bibfield{author}{\bibinfo{person}{V. Vapnik} {and} \bibinfo{person}{R.
  Izmailov}.} \bibinfo{year}{2015}\natexlab{}.
\newblock \showarticletitle{Learning using privileged information: Similarity
  control and knowledge transfer}.
\newblock \bibinfo{journal}{\emph{Journal of Machine Learning Research}}.
\newblock


\bibitem[\protect\citeauthoryear{Vapnik and Vashist}{Vapnik and
  Vashist}{2009}]%
        {vapnik2009new}
\bibfield{author}{\bibinfo{person}{V. Vapnik} {and} \bibinfo{person}{A.
  Vashist}.} \bibinfo{year}{2009}\natexlab{}.
\newblock \showarticletitle{A new learning paradigm: Learning using privileged
  information}.
\newblock \bibinfo{journal}{\emph{Neural Networks}}.
\newblock


\bibitem[\protect\citeauthoryear{Walls, Learned-Miller, and Levine}{Walls
  et~al\mbox{.}}{2011}]%
        {walls2011forensic}
\bibfield{author}{\bibinfo{person}{R.~J. Walls}, \bibinfo{person}{E.~G.
  Learned-Miller}, {and} \bibinfo{person}{B.~N. Levine}.}
  \bibinfo{year}{2011}\natexlab{}.
\newblock \showarticletitle{Forensic triage for mobile phones with DEC0DE}. In
  \bibinfo{booktitle}{\emph{USENIX Security}}.
\newblock


\bibitem[\protect\citeauthoryear{Wang and Ji}{Wang and Ji}{2015}]%
        {wang2015classifier}
\bibfield{author}{\bibinfo{person}{Z. Wang} {and} \bibinfo{person}{Q. Ji}.}
  \bibinfo{year}{2015}\natexlab{}.
\newblock \showarticletitle{Classifier learning with hidden information}. In
  \bibinfo{booktitle}{\emph{IEEE Computer Vision and Pattern Recognition}}.
\newblock


\bibitem[\protect\citeauthoryear{Wolf, Hassner, and Taigman}{Wolf
  et~al\mbox{.}}{2011}]%
        {wolf2011effective}
\bibfield{author}{\bibinfo{person}{L. Wolf}, \bibinfo{person}{T. Hassner},
  {and} \bibinfo{person}{Y. Taigman}.} \bibinfo{year}{2011}\natexlab{}.
\newblock \showarticletitle{Effective unconstrained face recognition by
  combining multiple descriptors and learned background statistics}.
\newblock \bibinfo{journal}{\emph{Pattern Analysis and Machine Intelligence}}.
\newblock


\bibitem[\protect\citeauthoryear{Yadav et~al\mbox{.}}{Yadav
  et~al\mbox{.}}{2010}]%
        {yadav2010detecting}
\bibfield{author}{\bibinfo{person}{S. Yadav} {et~al\mbox{.}}}
  \bibinfo{year}{2010}\natexlab{}.
\newblock \showarticletitle{Detecting algorithmically generated malicious
  domain names}. In \bibinfo{booktitle}{\emph{Internet Measurement
  Conference}}.
\newblock


\bibitem[\protect\citeauthoryear{Zou, Kesidis, and Miller}{Zou
  et~al\mbox{.}}{2011}]%
        {zou2011flow}
\bibfield{author}{\bibinfo{person}{G. Zou}, \bibinfo{person}{G. Kesidis}, {and}
  \bibinfo{person}{D.~J. Miller}.} \bibinfo{year}{2011}\natexlab{}.
\newblock \showarticletitle{A flow classifier with tamper-resistant features
  and an evaluation of its portability to new domains}.
\newblock \bibinfo{journal}{\emph{JSAC}}.
\newblock


\end{thebibliography}

\newpage
\appendix{}
\section{Model Influence Optimization}
\label{sec:appendixA}
In this Appendix, we present the formulation of model influence approach  introduced in Section~\ref{sec:modelInfluence}  (Section~\ref{sec:appendixA-formulation}) and its implementation in MATLAB (Section~\ref{sec:appendix-implementation}) in order to realize the paradigm in detection systems. 
\subsection{Model Influence Formulation}
\label{sec:appendixA-formulation}
\setcounter{equation}{0}
We can formally divide the feature space into two spaces at training time. Given $L$ standard vectors $\vec{x_1},\ldots,\vec{x_L}$ and $L$ privileged vectors $\vec{x_1^*},\ldots,\vec{x_L^*}$ with a target class $y=\{+1,-1\}$, where $\vec{x_i} \in \mathbb{R}^N$ and $\vec{x_i^*} \in \mathbb{R}^M$ for all $ i=1,\ldots,L$. The kernels $K(\vec{x_i},\vec{x_j})$ and $K^*(\vec{x_i^*},\vec{x_j^*})$ are selected along with positive parameters $\kappa$, and $\gamma$.  Our goal is finding a detection model $f \rightarrow \mathbf{x}^{s} : y$. The optimization problem is formulated as~\cite{vapnik2015learning}:
\begin{equation}
\label{LUPI-SVM2}
\left\{
\begin{array}{l}
\displaystyle\sum_{i=1}^{L}\alpha_i-\displaystyle\frac{1}{2}\displaystyle
\sum_{i,j=1}^{L}y_iy_j\alpha_i\alpha_jK(\vec{x_i},\vec{x_j})\\
-\displaystyle\frac{\gamma}{2}\displaystyle \sum_{i,j=1}^{L}y_iy_j
(\alpha_i-\delta_i)(\alpha_j-\delta_j)K^*(\vec{x_i^*},\vec{x_j^*})
\to{\rm max}\\[1em]
\displaystyle\sum_{i=1}^{L}\alpha_iy_i=0,\;\;\displaystyle\sum_{i=1}^{L}
\delta_i y_i =0\\[1em]
0\leq\alpha_i\leq\kappa C_i,\;\; 0\leq\delta_i\leq C_i,\;\;
i=1,\ldots,L
\end{array}
\right.
\end{equation}
\noindent The detection rule $f$ for vector $\vec{z}$ is defined as:
\begin{equation}\label{LUPI-SVM2-Decision-rule}
f(\vec{z})={\rm
sign}\left(\sum_{i=1}^Ly_i\alpha_iK(\vec{x_i},\vec{z})+B\right)
\end{equation}
where to compute $B$, we first derive the Lagrangian of~(\ref{LUPI-SVM2}):
\begin{align}
\label{LUPI-SVM2-Lagrangian}
{\mathcal L}(\vec{\alpha},\vec{\beta},\vec{\phi},\vec{\lambda},\vec{\mu},\vec{\nu},\vec{\rho})&=\displaystyle\sum_{i=1}^{L}\alpha_i-  \displaystyle\frac{1}{2}\displaystyle\nonumber \sum_{i,j=1}^{L}y_iy_j\alpha_i\alpha_jK(\vec{x_i},\vec{x_j}) \\ \nonumber 
&-\displaystyle\frac{\gamma}{2}\displaystyle \sum_{i,j=1}^{L}y_iy_j (\alpha_i-\delta_i)(\alpha_j-\delta_j)K^*(\vec{x_i^*},\vec{x_j^*})\\ \nonumber 
&+\phi_1\displaystyle\sum_{i=1}^{L}\alpha_iy_i+\phi_2\displaystyle\sum_{i=1}^{L}\delta_iy_i+\displaystyle\sum_{i=1}^{L}\lambda_i\alpha_i\\ \nonumber 
&+\displaystyle\sum_{i=1}^{L}\mu_i(\kappa C_i-\alpha_i)+
\displaystyle\sum_{i=1}^{L}\nu_i\delta_i\\ 
&+\displaystyle\sum_{i=1}^{L}\rho_i(C_i-\delta_i)
\end{align}
\noindent with Karush-Kuhn-Tucker (KKT) conditions (for each $i=1,\ldots,L$), we rewrite
\begin{align}
\label{LUPI-SVM2-KKT}
\displaystyle\frac{\partial{\mathcal
L}}{\partial\alpha_i}&=-K(\vec{x_i},\vec{x_i})\alpha_i -\gamma K^*(\vec{x_i^*},\vec{x_i}^*)\alpha_i\nonumber+ K^*(\vec{x_i^*},\vec{x_i^*})\gamma\delta_i\\\nonumber
& -\displaystyle\sum_{k\neq i}K(\vec{x_i},\vec{x_k})y_iy_k\alpha_k-\gamma\displaystyle\sum_{k\neq i}K^*(\vec{x_i^*},\vec{x_k^*})y_iy_k(\alpha_k-\delta_k) \\\nonumber
&+1+\phi_1 y_i+\lambda_i-\mu_i=0 \\  \nonumber 
\displaystyle\frac{\partial{\mathcal L}}{\partial\delta_i}&=-K^*(\vec{x_i^*},\vec{x_i^*})\gamma\delta_i + K^*(\vec{x_i^*},\vec{x_i^*})\gamma\alpha_i\\ \nonumber
& +\displaystyle\sum_{k\neq i}K(\vec{x_i},\vec{x_k})y_iy_k\gamma(\alpha_k-\delta_k)\\
& +\phi_2 y_i+\nu_i-\rho_i=0 
\end{align}
where
\begin{align}
&\lambda_i\geq 0, ~~\mu_i\geq 0, ~~\nu_i\geq 0,  ~~\rho_i\geq 0,\\ \nonumber
&\lambda_i\alpha_i=0, ~~\mu_i(C_i-\alpha_i)=0,\\ \nonumber
&\nu_i\delta_i=0, ~~\rho_i(C_i-\delta_i)=0,\\ \nonumber
&\displaystyle\sum_{i=1}^L\alpha_i y_i=0, ~~\displaystyle\sum_{i=1}^L\delta_i y_i=0 \nonumber
\end{align}
We denote for $i=1,\ldots,L$
\begin{align}
\label{LUPI-SVM2-Ff}
F_i&= \displaystyle\sum_{k=1}^{L}K(\vec{x_i},\vec{x_k})y_k\alpha_k,\\ \nonumber
f_i&= \displaystyle\sum_{k=1}^{L}K^*(\vec{x_i^*},\vec{x_k^*})y_k(\alpha_k-\delta_k)\\\nonumber
\end{align}
and rewrite (\ref{LUPI-SVM2-KKT}) in the form
\begin{equation}\label{LUPI-SVM2-KKT-simplified}
\left\{
\begin{array}{l}
\displaystyle\frac{\partial{\mathcal
L}}{\partial\alpha_i}=-y_iF_i-\gamma y_i
f_i+1+\phi_1 y_i+\lambda_i-\mu_i=0\\[1em]
\displaystyle\frac{\partial{\mathcal L}}{\partial\delta_i}=\gamma
y_if_i+\phi_2 y_y+\nu_i-\rho_i=0\\[1em]
\lambda_i\geq 0,\;\;\;\mu_i\geq 0,\;\;\; \nu_i\geq 0,
\;\;\rho_i\geq 0,\\[1em]
\lambda_i\alpha_i=0, \;\mu_i(C_i-\alpha_i)=0,
\;\;\nu_i\delta_i=0, \;\rho_i(C_i-\delta_i)=0\\
\displaystyle\sum_{i=1}^L\alpha_i y_i=0 ,\;\;
\displaystyle\sum_{i=1}^L\delta_i y_i=0
\end{array}
\right.
\end{equation}
The first equation in (\ref{LUPI-SVM2-KKT-simplified}) implies
\begin{equation}\label{LUPI-SVM2-phi}
\phi_1 =-y_j(1-y_jF_j-\gamma y_j f_j+\lambda_j-\mu_j)
\end{equation}
for all $j$. If $j$ is selected such that $0<\alpha_j<\kappa C_j$ and $0<\delta_j<C_j$, then (\ref{LUPI-SVM2-KKT-simplified}) implies $\lambda_j=\mu_j=\nu_j=\rho_j=0$ and (\ref{LUPI-SVM2-phi}) has the following form
\begin{align*}
\label{LUPI-SVM-phi-simplified}
\phi_1 &=-y_j(1-y_jF_j-\gamma y_j f_j)\\
\phi_1&=-y_j(( 1-\sum_{i=1}^L
y_iy_jK(\vec{x_i},\vec{x_j})(\alpha_i)\\
&-\gamma\sum_{i=1}^Ly_iy_jK^*(\vec{x_i^*},\vec{x_j^*})(\alpha_i-\delta_i))
\end{align*}
Therefore, $B$ is computed as $B=-\phi_1$:
\begin{align}
B&=y_j(1-\sum_{i=1}^L\nonumber
y_iy_jK(\vec{x_i},\vec{x_j})(\alpha_i)\\ 
&-\gamma\sum_{i=1}^Ly_iy_jK^*(\vec{x_i^*},\vec{x_j^*})(\alpha_i-\delta_i))
\end{align}
where $j$ is such that $0<\alpha_j<\kappa C_j$ and $0<\delta_j<C_j$.\\

\subsection{Model Influence Implementation}
\label{sec:appendix-implementation}
We present implementation of model influence by solving its quadratic programming problem using MATLAB \texttt{quadprog} function provided by the optimization toolbox. Other equivalent functions in R~\cite{turlach2007quadprog} or similar software can be easily adapted. 

MATLAB function ${\rm \texttt{quadprog}}(H,\vec{f},A,\vec{b},Aeq,\vec{beq},\vec{lb},\vec{ub})$ solves the quadratic programming problem in the form as follows:
\begin{equation}\label{LUPI-SVM2-Matlab}
\left\{
\begin{array}{l}
\displaystyle\frac{1}{2}\vec{z}^TH\vec{z}+f^T\vec{z}\to{\rm min}\\[1em]
A \cdot \vec{z}\leq\vec{b}\\
Aeq \cdot \vec{z}=\vec{beq}\\
\vec{lb}\leq\vec{z}\leq\vec{ub}
\end{array}
\right.
\end{equation}
\noindent Here, $H, A, Aeq$  are matrices, and $\vec{f}, \vec{b}, \vec{beq}, \vec{lb}, \vec{ub}$ are vectors. $\vec{z}$ is defined as $\vec{z}=(\alpha_1,\ldots,\alpha_L,\delta_1,\ldots,\delta_L)\in R^{2L}$. We now rewrite~(\ref{LUPI-SVM2}) in the form of (\ref{LUPI-SVM2-Matlab}).
$$
\displaystyle\frac{1}{2}\vec{z}^TH\vec{z}+f^T\vec{z}\to{\rm min}
$$
where
$$f=(-1_1,-1_2,\ldots,-1_L,0_{L+1},\ldots,0_{2L})$$
and
$$
H_{ij}=\left( \begin{array}{ll}H^{11} & H^{12} \\
H^{12} & H^{22}\end{array}\right)
$$
where, for each pair $i,j=1,\ldots,L$,
\begin{align*}
H^{11}_{ij}&=K(\vec{x_i},\vec{x_j}) y_iy_j +\gamma K^*(\vec{x_i^*},\vec{x_j^*})y_iy_j ,\;\; \\
H^{12}_{ij}&=-\gamma K^*(\vec{x_i^*},\vec{x_j^*})y_iy_j,\;\;\\
H^{22}_{ij}&=+\gamma K^*(\vec{x_i^*},\vec{x_j^*})y_iy_j
\end{align*}
The second line of~(\ref{LUPI-SVM2-Matlab}) is absent. The third line of~(\ref{LUPI-SVM2-Matlab}) corresponds to the second line of~(\ref{LUPI-SVM2}) when written as
$$
Aeq \cdot \vec{z}=\vec{beq}
$$
where
$$
Aeq= \left(
\begin{array}{cccccccc}
y_1 & y_2 & \cdots & y_L & 0 &   0 & \cdots &  0  \\
  0 &   0 & \cdots &  0  & y_1 & y_2 & \cdots & y_L
\end{array}
\right),\;\;\ 
$$

$$
\vec{beq}=\left(\begin{array}{c}0 \\
0\end{array}\right)
$$
The fourth line of~(\ref{LUPI-SVM2-Matlab}) corresponds to the third line of~(\ref{LUPI-SVM2}) when written as
$$
\vec{lb}\leq\vec{z}\leq\vec{ub}
$$
where
$$
\begin{array}{l}
\vec{lb}=(0_1,0_2,\ldots,0_L,0_{L+1},\ldots,0_{2L}),\;\\
\vec{ub}=(\kappa C_1,\kappa C_2,\ldots,\kappa
C_L,C_1,C_2,\ldots,C_L)
\end{array}
$$

After all variables $(H,\vec{f},A,\vec{b},Aeq,\vec{beq},\vec{lb},\vec{ub})$ are defined, optimization toolbox guide~\cite{matlabQuadprog} can be used to select \texttt{quadprog()} function options such as an optimization algorithm and maximum number of iterations. Then, output of the function can be used in detection function $f$ for a new sample $\vec{z}$ to make predictions as follows:

\begin{equation}\label{LUPI-SVM2-Decision-ruleFinal}
f(\vec{z})={\rm
sign}\left(\sum_{i=1}^Ly_i\alpha_iK(\vec{x_i},\vec{z})+B\right)
\end{equation}

\section{Details of Detection Systems}
\label{sec:AppendixC}
\setcounter{table}{0}
\setcounter{figure}{0}
In this Appendix, we detail the standard and privileged features of fast-flux bot and malware traffic detection systems introduced in Section~\ref{sec:exp-systems}.  Table~\ref{table:ffdetector} presents feature categories and definitions of fast-flux bot detector obtained from recent works~\cite{huang2010fast,hsu2010fast,yadav2010detecting,passerini2008fluxor}, and Table~\ref{table:longfeatures} presents the features of malware traffic detector obtained from recent works~\cite{miller2012sequential,celik2011salting, zou2011flow, berkay2015malware}. The interested reader can refer to the references for the motivation of the feature selection. 

\begin{table*}[h!]
\centering
\resizebox{\textwidth}{!}{
\small{
  \renewcommand{\arraystretch}{1.2}
  \begin{tabular}[width=\textwidth]{p{1.6cm} p{6cm} p{7cm} p{2.5cm} }
  \hline\bfseries Category&\bfseries Definition  &\bfseries Feature dependency &\bfseries Feature type \\ \hline
   DNS Answer
  & Number of unique A records\newline Number of NS records
  & DNS packet analysis 
  & standard set \\\hline    
  ~ \newline  Timing
  &  Network delay ($\mu$ and $\sigma$) \newline Processing delay ($\mu$ and $\sigma$) \newline Document fetch delay ($\mu$ and $\sigma$)
  &  ~ \newline HTTP requests 
  & ~ \newline standard set\\\hline 
  ~ \newline Domain name 
  & Edit distance\newline Kullback-Leibler divergence (unigrams and bigrams)\newline Jaccard similarity (unigrams and bigrams)
  &~\newline  Whitelist of benign domain names   
  &~ \newline  privileged set \\\hline
  ~ \newline Spatial
  & Time zone entropy of A records \newline Time zone entropy of NS records \newline Minimal service distances ($\mu$ and $\sigma$)
  &  IP coordinate database lookup (external source)
  &   privileged set \\\hline
  Network
  & Number of distinct autonomous systems  \newline Number of distinct networks
  & WHOIS processing (external source) 
  & privileged set  \\\hline    
  \end{tabular}
  }
  }
\caption{Fast-flux bot detection system standard and privileged feature descriptions ($\mu$ is mean and $\sigma$ is std. dev.).}
\label{table:ffdetector}
\end{table*}

\begin{table*}[h!]
\resizebox{\textwidth}{!}{
\small{
\renewcommand{\arraystretch}{1.2} 
\begin{tabular}[width=\textwidth]{p{1.8cm}| p{5.3cm}|p{7.4cm}| p{1.5cm}}\hline  
\multicolumn{1}{l}{\textbf{Abbreviation}} &
\multicolumn{1}{c}{\textbf{Definition}} &
\multicolumn{1}{c}{\textbf{Properties}}&
\multicolumn{1}{l}{\textbf{Feature type}}\\ \hline 
 cnt-data-pkt & The count of all the packets with at least a byte of TCP data payload & -TCP length is observed \newline  -Client to server  & standard set\\ 
min-data-size & The minimum payload size observed & -TCP length observed\newline -Client to server\newline -0 if there are no packets  & standard set\\
avg-data-size & Data bytes divided by the total number of packets&-TCP length observed\newline -Packets with payload observed \newline -Server to client \newline -0 if there are no packets   & standard set\\
init-win-bytes & The total number of bytes sent in initial window & -Retransmitted packets not counted\newline -Client to server \& server to client \newline -0 if no ACK observed\newline -Frame length calculated  & standard set  \\
RTT-samples & The total number of RTT samples  found & -Client to server & standard set \\
IP-bytes-median & Median of total IP packets & -IP length calculated\newline -Client to server  & standard set\\
frame-bytes-var & Variance of bytes in Ethernet packets & -Frame length calculated\newline -Client to server   & standard set\\
IP-ratio & Ratio between the maximum packet size and minimum packet size & -IP length calculated\newline -Client to server \& server to client\newline -1 If a packet observed, and if no packets are observed 0 is reported  & standard set\\
pushed-data-pkts & The count of all the packets seen with the PUSH set in TCP header & -Client to server \& server to client & standard set\\
goodput & Total number of frame bytes divided by the differences between last packet time and first packet time & -Frame length calculated\newline -Client to server\newline -Retransmitted bytes not counted  & standard set\\
duration & Total connection time& Time difference between the last packet and first packet (SYN flag is seen from destination) & privileged set\\
min-IAT & Minimum packet inter-arrival time for all packets of the flow & -Client to server \& server to client  & privileged set\\
urgent-data-pkts  & The total number of packets with the URG bit turned on in the TCP header  & -Client to server \& server to client & privileged set\\
src-port & Source port number&-Undecoded & privileged set\\ 
dst-port & Destination port number&-Undecoded & privileged set\\ 
payload-info& Byte frequency distributions&-If not HTTPS at training time\newline-If payloads are available\newline-Client to server \& server to client & privileged set\\
\end{tabular}
}
}
\caption{Malware traffic detection standard and privileged feature descriptions.} 
\label{table:longfeatures} 
\end{table*}

\end{document}